\documentclass[twocolumn]{aastex62}

\usepackage{sidecap}
\usepackage{color}

\graphicspath{{./}{figures/}}

\shorttitle{The UV and X-Ray Evolution of K Stars}
\shortauthors{Richey-Yowell et al. (2018)}

\begin{document}
\title{HAZMAT. V. The Ultraviolet and X-ray Evolution of K Stars}

\author{Tyler Richey-Yowell}
\affil{School of Earth and Space Exploration, Arizona State University, Tempe, AZ 85281, USA}
\email{tricheyy@asu.edu}

\author{Evgenya L. Shkolnik}
\affil{School of Earth and Space Exploration, Arizona State University, Tempe, AZ 85281, USA}

\author{Adam C. Schneider}
\affil{School of Earth and Space Exploration, Arizona State University, Tempe, AZ 85281, USA}

\author{Ella Osby}
\affil{School of Earth and Space Exploration, Arizona State University, Tempe, AZ 85281, USA}

\author{Travis Barman}
\affil{Lunar and Planetary Laboratory, University of Arizona, Tucson, AZ 85721, USA}

\author{Victoria S. Meadows}
\affil{NASA Astrobiology Institute, Virtual Planetary Laboratory, University of Washington, Seattle, WA 85215, USA}

\begin{abstract}

Knowing the high-energy radiation environment of a star over a planet's formation and evolutionary period is critical in determining if that planet is potentially habitable and if any biosignatures could be detected, as UV radiation can severely change or destroy a planet's atmosphere. Current efforts for finding a potentially habitable planet are focused on M stars, yet K stars may offer more habitable conditions due to decreased stellar activity and more distant and wider habitable zones (HZ). While M star activity evolution has been observed photometrically and spectroscopically, there has been no dedicated investigation of K-star UV evolution. We present the first comprehensive study of the near-UV, far-UV, and X-ray evolution of K stars. We used members of young moving groups and clusters ranging in age from 10 -- 625 Myr combined with field stars and their archived \textit{GALEX} UV and \textit{ROSAT} X-ray data to determine how the UV and X-ray radiation evolve. We find that the UV and X-ray flux incident on a HZ planet is 5 - 50 times lower than that of HZ planets around early-M stars and 50 - 1000 times lower than those around late-M stars, due to both an intrinsic decrease in K dwarf stellar activity occurring earlier than for M dwarfs and the more distant location of the K dwarf HZ. 

\end{abstract}

\accepted{December 21, 2018}
\submitjournal{ApJ}

\section{Introduction}\label{sec:intro}

\begin{deluxetable*}{c c c c c c c c}[th!]
\tablecaption{\small{Mass estimates based on the spectral type to effective temperature transformations of \citet{Pecaut2013} and the evolutionary models of \citet{Baraffe2015}. } \label{tab:massest}}
\tablecolumns{8}
\tablewidth{0pt}
\tablehead{
\colhead{SpT} &
\colhead{TW Hya} &
\colhead{$\beta$ Pic} & 
\colhead{Tuc-Hor} & 
\colhead{AB Dor} &
\colhead{UMa} & 
\colhead{Hyades} & 
\colhead{Field} \\
\colhead{} & \colhead{10 Myr} & \colhead{24 Myr} & \colhead{45 Myr} & \colhead{149 Myr} & \colhead{300 Myr} & \colhead{625 Myr} & \colhead{5 Gyr}
}
\startdata
K0 & 1.30 & 0.98 & 0.90 & 0.93 & 0.92 & 0.92 & 0.89\\
K1 & 1.26 & 0.96 & 0.88 & 0.89 & 0.89 & 0.89 & 0.86\\
K2 & 1.20 & 0.93 & 0.86 & 0.86 & 0.86 & 0.86 & 0.83\\
K3 & 1.10 & 0.89 & 0.83 & 0.80 & 0.80 & 0.80 & 0.79\\
K4 & 0.98 & 0.83 & 0.80 & 0.75 & 0.75 & 0.75 & 0.74\\
K5 & 0.90 & 0.78 & 0.77 & 0.71 & 0.71 & 0.71 & 0.70\\
K6 & 0.78 & 0.74 & 0.72 & 0.63 & 0.65 & 0.66 & 0.65\\
K7 & 0.75 & 0.72 & 0.69 & 0.59 & 0.62 & 0.62 & 0.61\\
K8 & 0.72 & 0.71 & 0.66 & 0.57 & 0.60 & 0.60 & 0.60\\
K9 & 0.68 & 0.68 & 0.63 & 0.54 & 0.57 & 0.57 & 0.57\\
M0 & 0.59 & 0.62 & 0.60 & 0.51 & 0.56 & 0.53 & 0.53\\
M1 & 0.49 & 0.52 & 0.52 & 0.45 & 0.50 & 0.47 & 0.47
\enddata
\end{deluxetable*}

Astronomers have discovered thousands of exoplanets, with an estimate for the occurrence rate of Earth-sized planets in the habitable zone (HZ) around GK stars of 22\% \citep{Petigura2013} from the Kepler mission statistics. The \textit{Transiting Exoplanet Survey Satellite} (\textit{TESS}) is expected to detect yet another 3,500 planetary candidates during its mission life \citep{Ricker2009, Ricker2014, barclay18, huang18}. While \textit{TESS} is expected to discover $\sim$700 new planets around K stars, none are expected to be in the HZ since \textit{TESS} is optimized to find planets with short orbital periods, excluding the longer orbital periods of habitable zone planets around K stars \citep{barclay18, huang18}. However, the \textit{PLAnetary Transits and Oscillations of stars} (\textit{PLATO}) mission, which will be optimized to find habitable zone planets around K dwarfs, is expected to discover $\sim$30 HZ K dwarf planets in the next decade (PLATO Definition Study Report, 2017). 

Further research is being conducted on what stellar characteristics increase the chances of HZ planets being in fact habitable \citep[e.g.][]{Cockell2016, Kaltenegger2017}. This would require a balance of environmental factors such as stellar lifetime, ultraviolet (UV) radiation, stellar winds, and HZ width. Currently, the most promising candidates for potentially habitable planets are thought to be around M stars (2,400 -- 3,700 K; 0.08 -- 0.45 M$_{\odot}$; conservative HZ width 0.1AU, \citealt{Kopparapu2013}); over half of all known HZ planets are around M dwarfs. Rocky planets around these stars are more easily detectable than around Sun-like stars due to favorable star-planet radius ratios, mass ratios, and shorter periods \citep{Charbonneau2007}. However, these stars are typically the most active of all spectral types and HZ planets around them are close-in and most likely tidally locked \citep[e.g.][]{Checlair2017, Barnes2017}. Planets around M stars that were initially expected to be likely habitable, such as Proxima Centauri b, are losing favorability due to excessive flaring (\citealt{2018ApJ...855L...2M}). M stars have been shown to have the largest levels of stellar activity during their pre-main sequence, which lasts through the duration of the formation of the planet and both its primordial and secondary atmospheres \citep{Schneider2018, Shkolnik2014}. Additionally, the large magnetic fields ($\sim$kG) of most M stars has been shown to potentially create magma oceans or partial melt the mantle of close-in planets, leading to increased volcanism and strong outgassing of greenhouse gases \citep{Kislyakova2017, Kislyakova2018}. K stars (3,700 -- 5,200K; 0.6 -- 0.9M$_{\odot}$), with conservative HZ width of 0.4AU, \citep{Kopparapu2013} may offer super-habitable conditions due to decreased stellar activity, lower magnetic field strengths, faster contraction onto the main sequence, and more distant and wider HZs. The highest potential of finding a truly habitable planet may lie with K stars.

The evolution of stellar activity as determined in the near-UV (NUV), far-UV (FUV), and X-ray wavelength regions has been studied in great detail for both GK stars \citep[e.g.][]{Findeisen2011} and M stars \citep[e.g.][]{stezler13, Shkolnik2014, Schneider2018} using archived \textit{Galaxy Evolution Explorer} (\textit{GALEX}; \citealt{Martin2004}) and \textit{R{\"O}ntgenSATellit} (\textit{ROSAT}; \citealt{rosat}) photometric data. However, a comprehensive investigation of the high-energy radiation environments of K-type stars has yet to be performed. The purpose of this study is to fill in this gap in our understanding of stellar evolution.

\begin{deluxetable*}{c c c c c c c c}[th!]
\tabletypesize{\footnotesize}
\tablecaption{\normalsize{Number of K stars in each group or cluster.} \label{tab:galex}}
\tablecolumns{8}
\tablewidth{0pt}
\tablehead{
\colhead{Group/Cluster} &
\colhead{Age} & 
\colhead{Input} & 
\colhead{Observed in NUV} &
\colhead{Detected in NUV} &
\colhead{Observed in FUV} &
\colhead{Detected in FUV} &
\colhead{Detected in X-ray} \\ 
\colhead{} & \colhead{[Myr]} & \colhead{Sample} & \colhead{by \textit{GALEX}} & \colhead{by \textit{GALEX}} & \colhead{by \textit{GALEX}} & \colhead{by \textit{GALEX}} & \colhead{by \textit{ROSAT}} 
}
\startdata
TWA  &  10 $\pm$ 3\tablenotemark{a} & 4 & 3 & 2 & 2 & 2 & 4 \\
Beta Pic  &  24 $\pm$ 3\tablenotemark{b}  & 30 & 21 & 19 & 16 & 15 & 27 \\
Columba  & 42$^{+6}_{-4}$\tablenotemark{a} & 14 & 11 & 10 & 10 & 5 & 4 \\
Carina  & 45$^{+11}_{-7}$\tablenotemark{a} & 3 & 1 & 1 & 1 & 1 & 3 \\
Tuc-Hor  &  45 $\pm$ 4\tablenotemark{a}  & 26 & 25 & 24 & 24 & 22 & 20 \\
AB Dor  &  149$^{+51}_{-19}$\tablenotemark{a}  & 12 & 11 & 6 & 10 & 10 & 12 \\
Ursa Major  & 414 $\pm$ 23\tablenotemark{c} & 12 & 7 & 3 & 7 & 7 & 8 \\
Praesepe  & 600\tablenotemark{d} & 133 & 52 & 49 & 62 & 17 & 0 \\
Hyades  &  625 $\pm$ 50\tablenotemark{e}  & 169 & 110 & 97 & 92 & 30 & 52 \\
Field  &  $\sim$5000  & 310 & 229 & 169 & 211 & 152 & 107 \\
\enddata
\tablerefs{
(a) \citet{Bell2015},
(b) \citet{Shkolnik2017},
(c) \citet{Jones15b, Jones17},
(d) \citet{Kraus2007},
(e) \citet{Perryman1997}
}
\end{deluxetable*}
\normalsize

\subsection{The Importance of UV Data}

The UV radiation incident on a planet is a critical factor in determining if that planet is potentially habitable and if any biosignatures could be detectable. UV radiation ionizes and photodissociates some of the most important molecules for the development of life (e.g. H$_2$O, H$_2$, O$_2$, CO$_2$, H$_2$S, and OCS) and signatures that would be indicative of life  (e.g. CH$_4$, NH$_3$, and N$_2$O) with potential for complete erosion of the atmosphere \citep[e.g.][]{Kasting1993, Lichtenegger2010, Segura2010, Hu2012}. UV radiation can be split into three regimes: the extreme-UV (EUV, 100-912 \AA), the far-UV (FUV, 912-1700 \AA), and the near-UV (NUV, 1700-3200 \AA) as defined in \citet{Linsky2015}. The EUV heats and inflates a planets upper-atmosphere, exacerbating its erosion \citep{Koskinen2010, Lammer2007}. Other than very limited data from the former \textit{Extreme Ultraviolet Explorer} (\textit{EUVE}) satellite, no information in this wavelength regime exists for stars other than the Sun. Any missions observing in the EUV will exclude most of the ionizing flux due to absorption by the ISM. These values are crucial as photochemical atmospheric models of HZ planets and abundance rate models require an input of EUV stellar fluxes, currently being interpolated from a limited number of FUV and X-ray observations. 

Realistic information regarding the evolution of the NUV, FUV, and EUV is of the utmost importance for using models to answer questions regarding planetary atmospheres. UV radiation, in particular the ratio of FUV/NUV, can impact the abundance rates and photochemistry of the atmospheres in the production of hazes in depleting atmospheres \citep{Zerkle2012, Arney2017} and ozone in oxidizing atmospheres \citep{Segura2003, Segura2005}, both of which drastically alter the planetary spectrum. This affects not only the composition, but also the detectability and stability of the atmospheres of these planets. For future missions such as the \textit{James Webb Space Telescope} (\textit{JWST}) that will focus on spectra, a planet with a hazy or depleted atmosphere will not make a good target. 

\begin{figure*}[h]
\centering
\includegraphics[width=.84\linewidth]{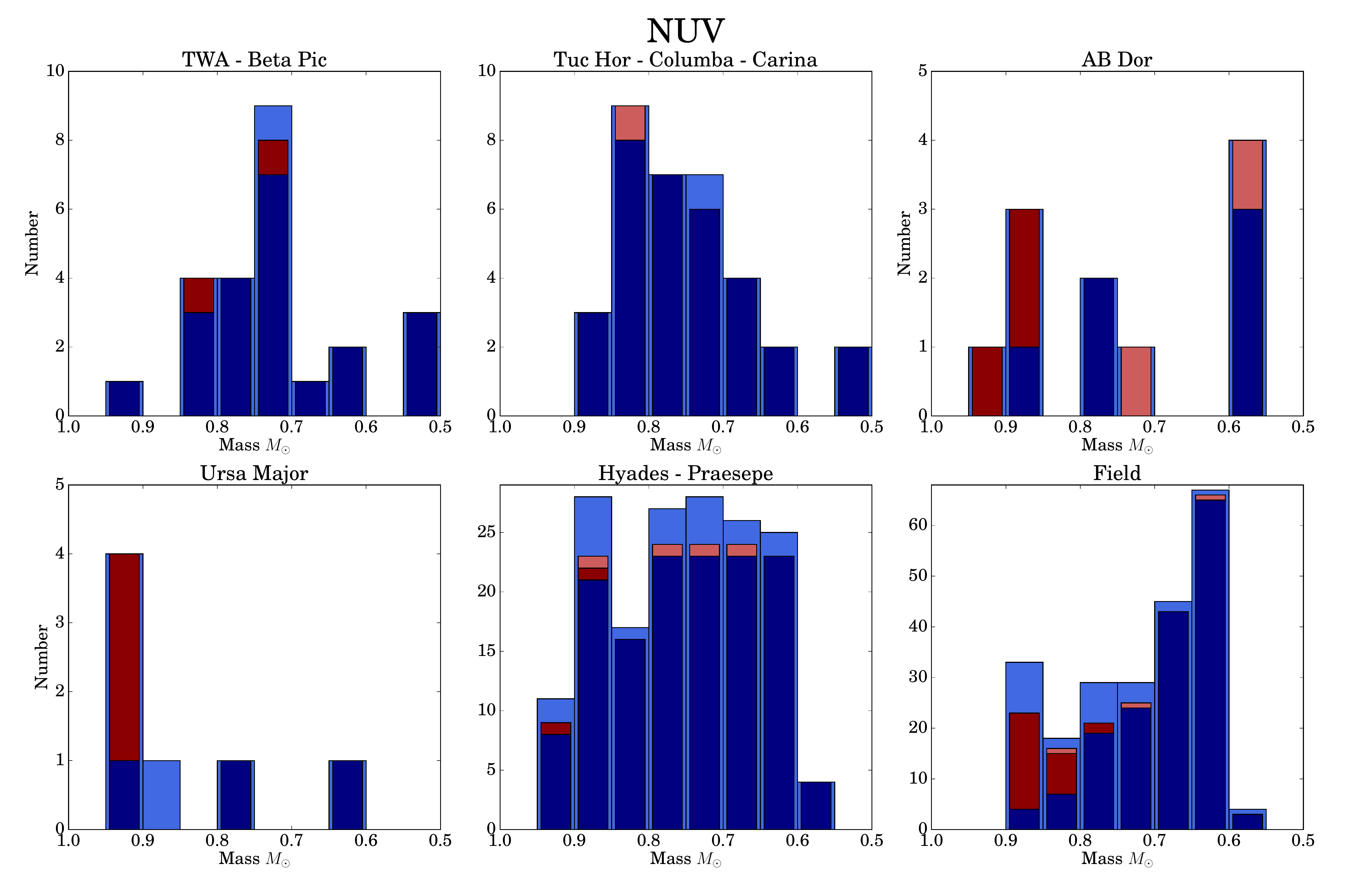}
\caption{Stacked histograms showing the mass distributions for the input and returned targets in the GALEX NUV for each age group. Light blue represents the input sample, dark blue are the detections with no photometric flags, dark red are the lower limits, and light red are the upper limits.  \label{fig:massdistributions_nuv}}
\vspace{10pt}
\end{figure*}

\begin{figure*}[h]
\centering
\includegraphics[width=.84\linewidth]{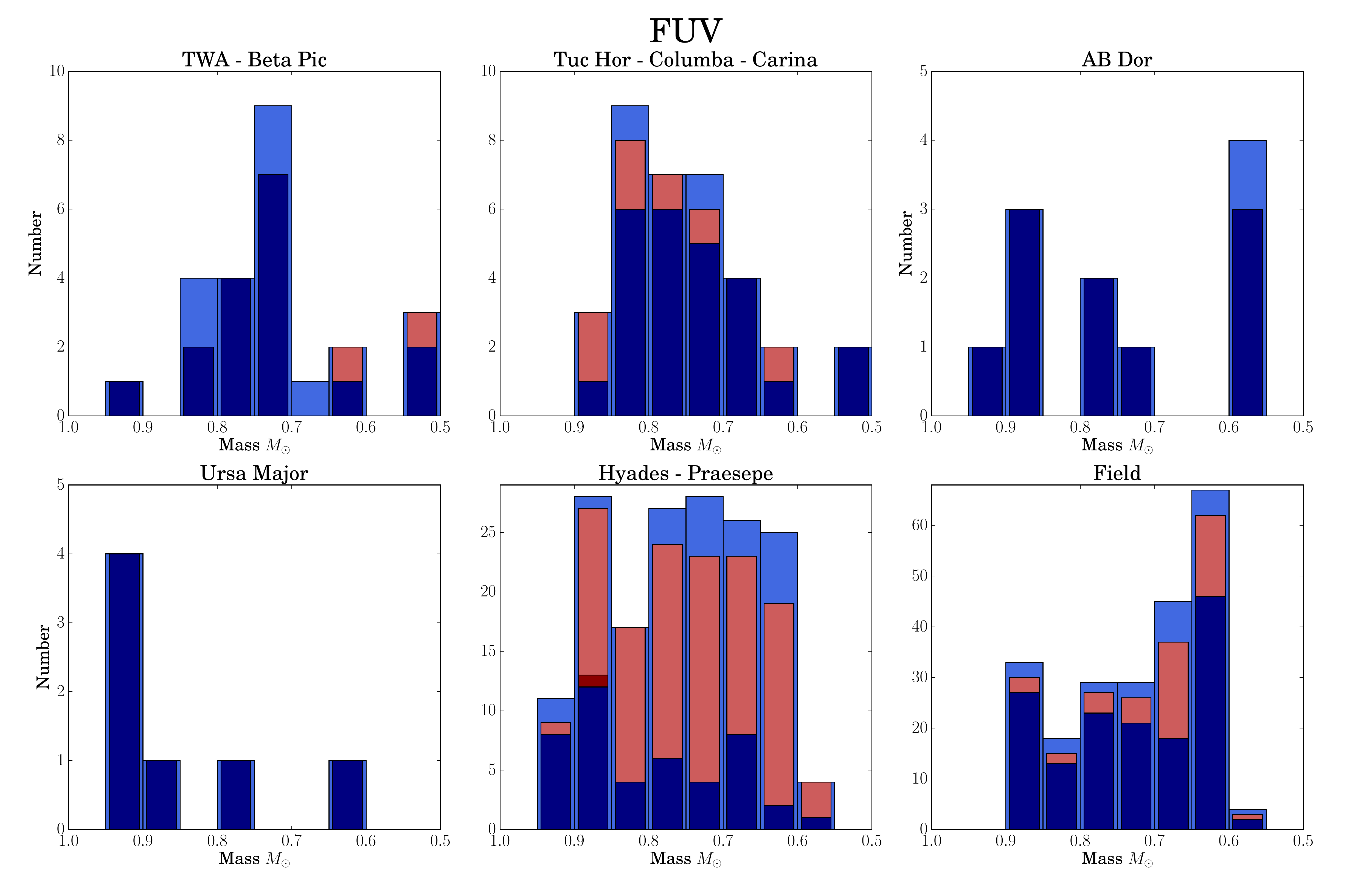}
\caption{Same as Figure \ref{fig:massdistributions_nuv} but for GALEX FUV targets.  \label{fig:massdistributions_fuv}}
\vspace{10pt}
\end{figure*}

\subsection{The K Star Advantage}\label{sec:kstaradvantage}

\begin{figure*}[th]
\centering
\includegraphics[width=0.8\linewidth]{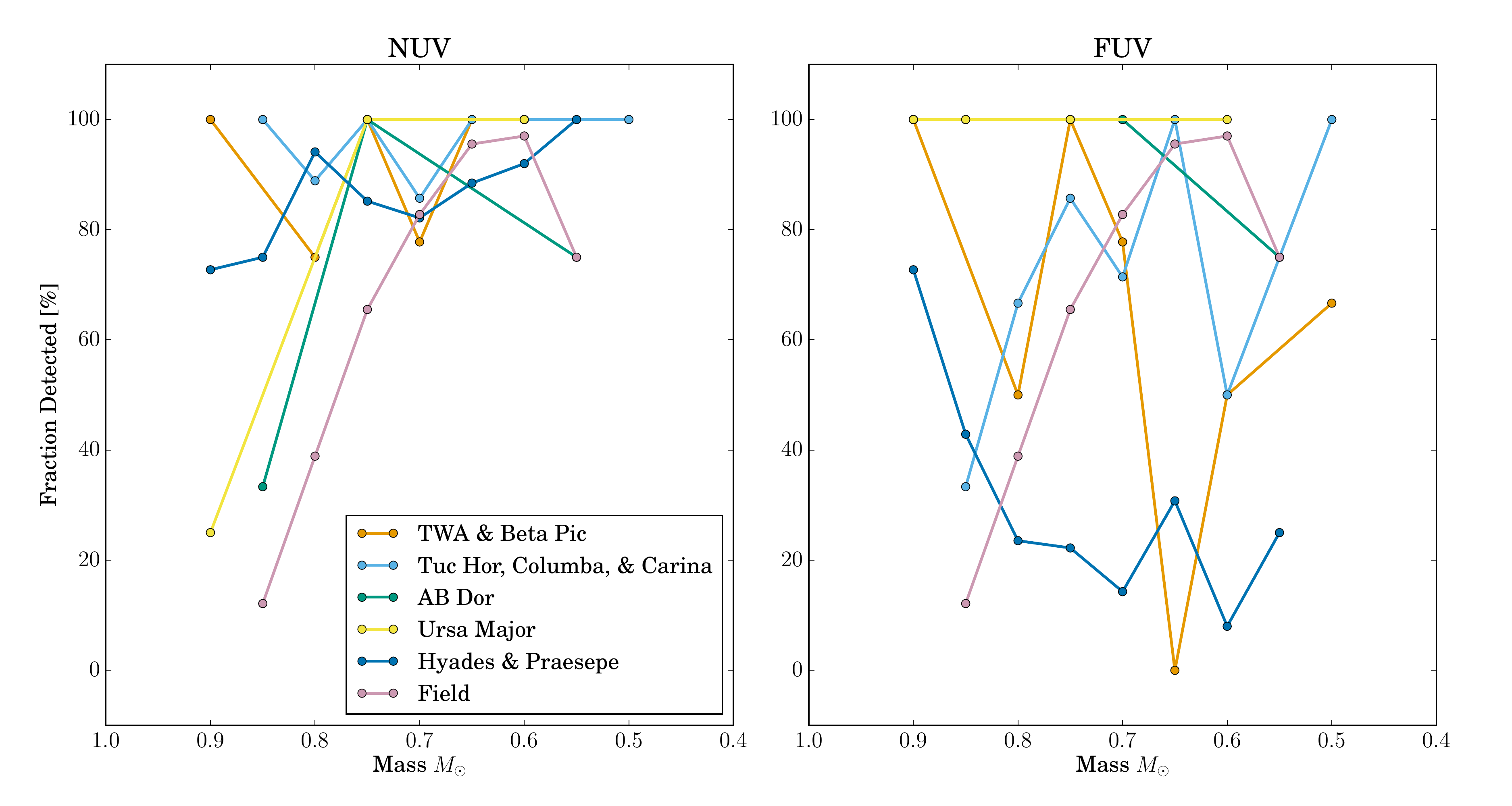}
\caption{Percentages of targets detected by \textit{GALEX} without any artifacts or flags compared to the input sample as a function of stellar mass. Groups of similar ages were combined to avoid issues due to small number statistics. In the NUV, higher-mass stars with a magnitude $<15$, i.e. in the non-linear regime, were considered instead as lower limits. The low return of Hyades and Praesepe members in the FUV is due to the clusters' further distances of 47pc and 177pc, respectively \citep{vanleeuwen}. \label{fig:mass_fractions}}
\end{figure*}

\begin{figure*}[h]
\centering
\includegraphics[width=0.85\linewidth]{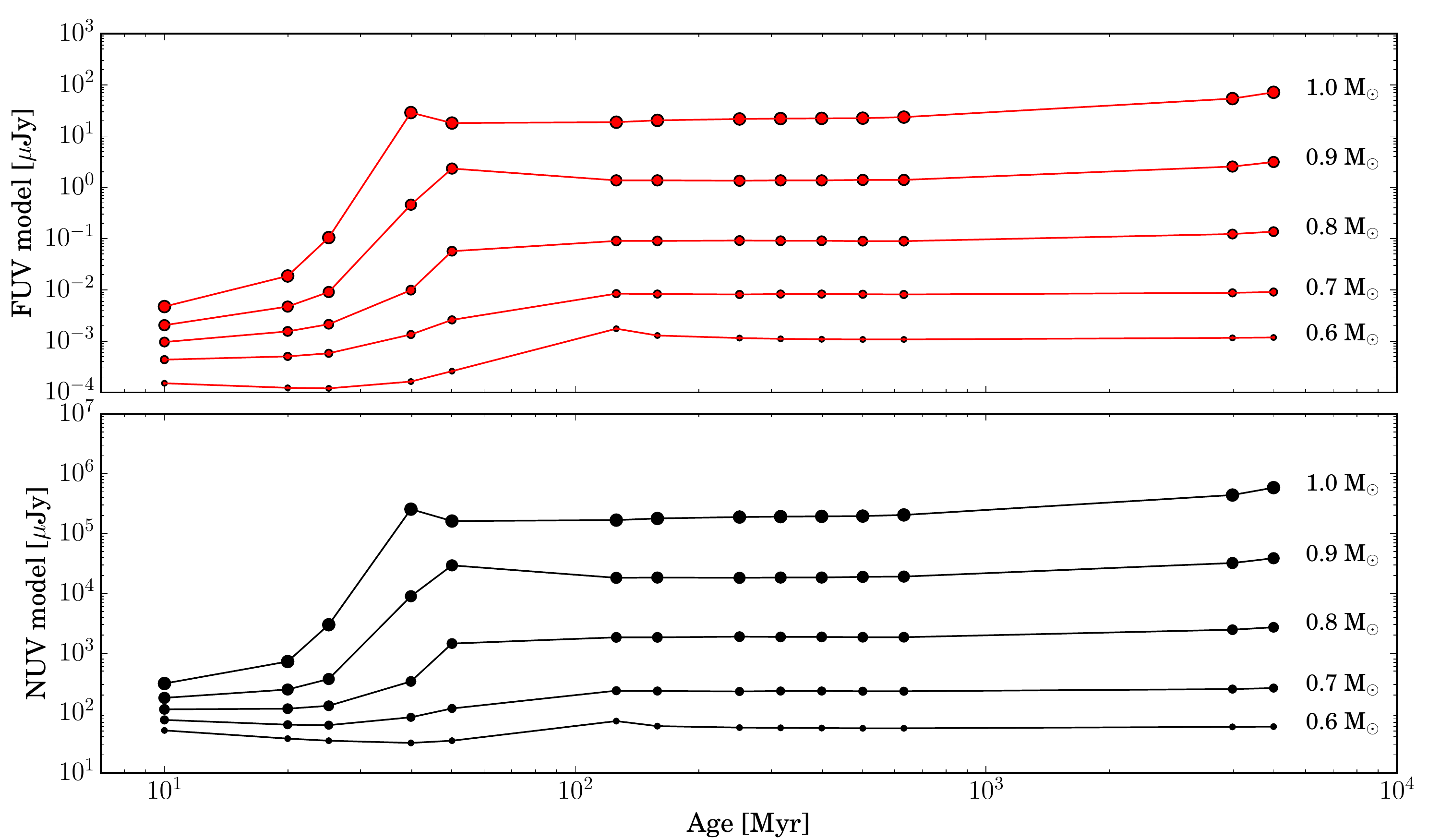}
\caption{Photospheric FUV and NUV flux densities calculated from the PHOENIX stellar atmosphere models as a function of age. Increasing point size represents increasing mass of the star from 0.6 -- 1.0 M$_{\odot}$. The 1.0 M$_{\odot}$ model is included for reference to the Sun. \label{fig:mvd_vs_age}}
\end{figure*}

\begin{figure*}[h]
\centering
\includegraphics[width=0.73\linewidth]{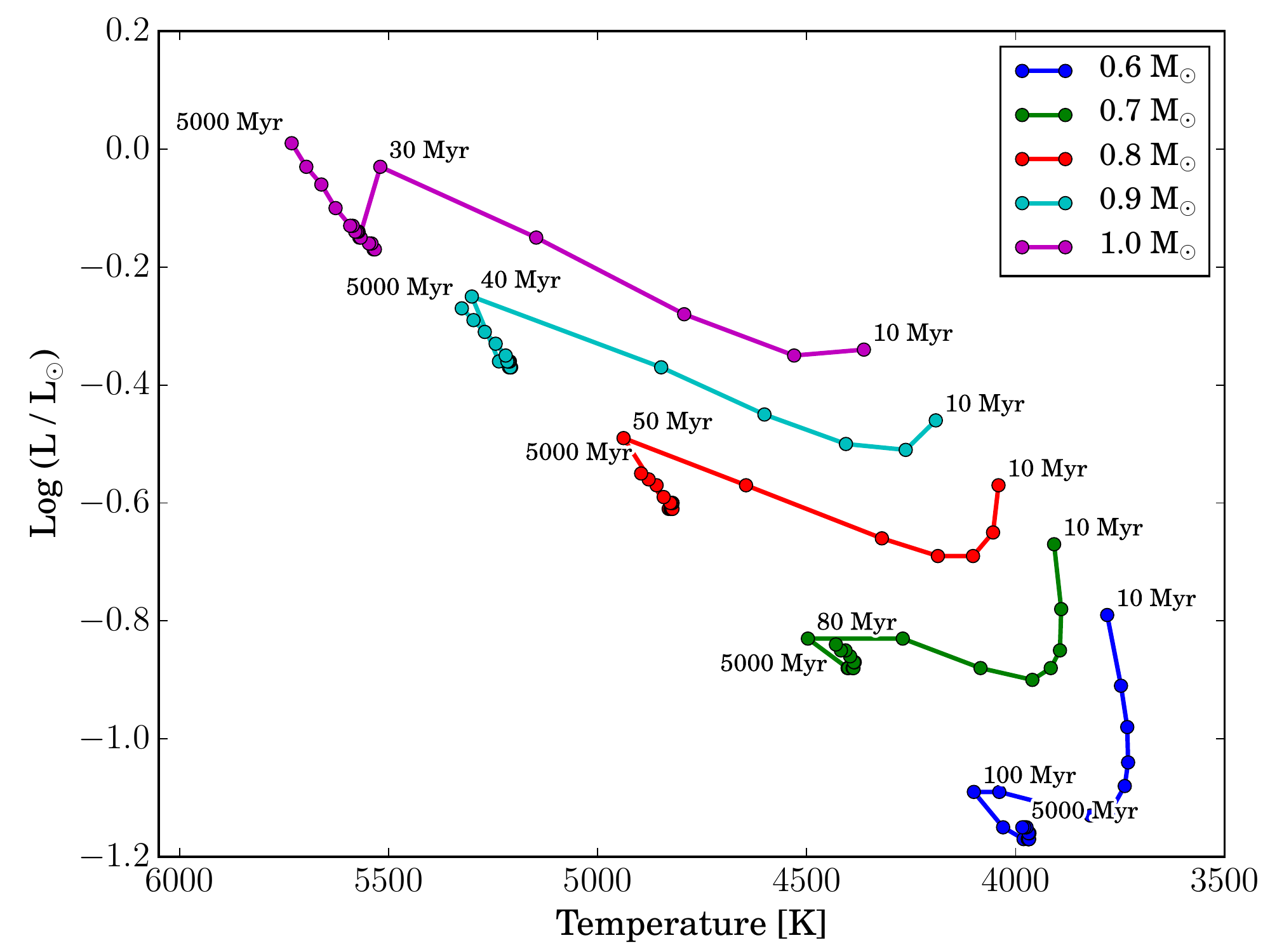}
\caption{H-R diagram showing the evolutionary tracks of stars from 0.6 -- 1.0 M$_{\odot}$ using the results of \citet{Baraffe2015}. Representative ages have been labeled to help guide the reader. Comparing this with Figure \ref{fig:mvd_vs_age}, it is clear that the drastic temperature change of pre-main sequence stars leads to a significant increase in the photospheric flux density, whereas the temperature (and thus the photospheric flux density) of main sequence stars is fairly constant.}  \label{fig:hr}
\vspace{10pt}
\end{figure*}

While current popularity resides among M stars as the most likely to host identifiable habitable planets, stellar variability, excess flaring and tidal locking in the HZ is cause for concern \citep[e.g.][]{Shields2016}. Early and mid-K spectral-type stars have been suggested by \citet{Heller2014} to be ``super-habitable'', i.e. the probability of life on a planet orbiting in the HZ of one of these super-habitable stars is higher than the probability of life on an Earth-sized planet orbiting in the HZ of a G2V star. \citet{Cuntz2016} analyzed several factors to determine the Habitable-Planetary-Real-Estate Parameter (HabPREP), which depends on the frequency of different stars, the rate of contraction onto the main sequence, the size and longevity of the HZ, UV and X-ray emission and flare frequency and power. They individually determined that the highest probability of having a habitable planet is in the early-K star regime. The longer lifetimes of the continuous HZ of these types of stars compared to F and G stars allow for life to tune their environment and develop a more favorable biosignature, whereas unmagnetized planets around M stars may lose their atmosphere within the first 100 Myr based on severe UV irradiance \citep{Airapetian2017}. \citet{Lingam2017} find that K6V stars around 0.67 M$_{\odot}$  would take the least amount of time for complex life to develop since increased oxygen levels from UV photolysis permit the emergence of complex life. \citet{Arney2018} have independently concluded that there is a similar “K star advantage” because a planet orbiting a K6V star has the potential to produce detectable CH$_4$ and O$_2$. However, the UV flux for most of these calculations was assumed to be constant throughout the timescale of the formation of life and could consequently lead to incorrect determinants. K star UV evolution is thus needed to determine the probability of planetary habitability. The highest potential of finding a truly habitable planet may lie with K stars.

\begin{figure*}[t]
\centering
\includegraphics[width=0.9\linewidth]{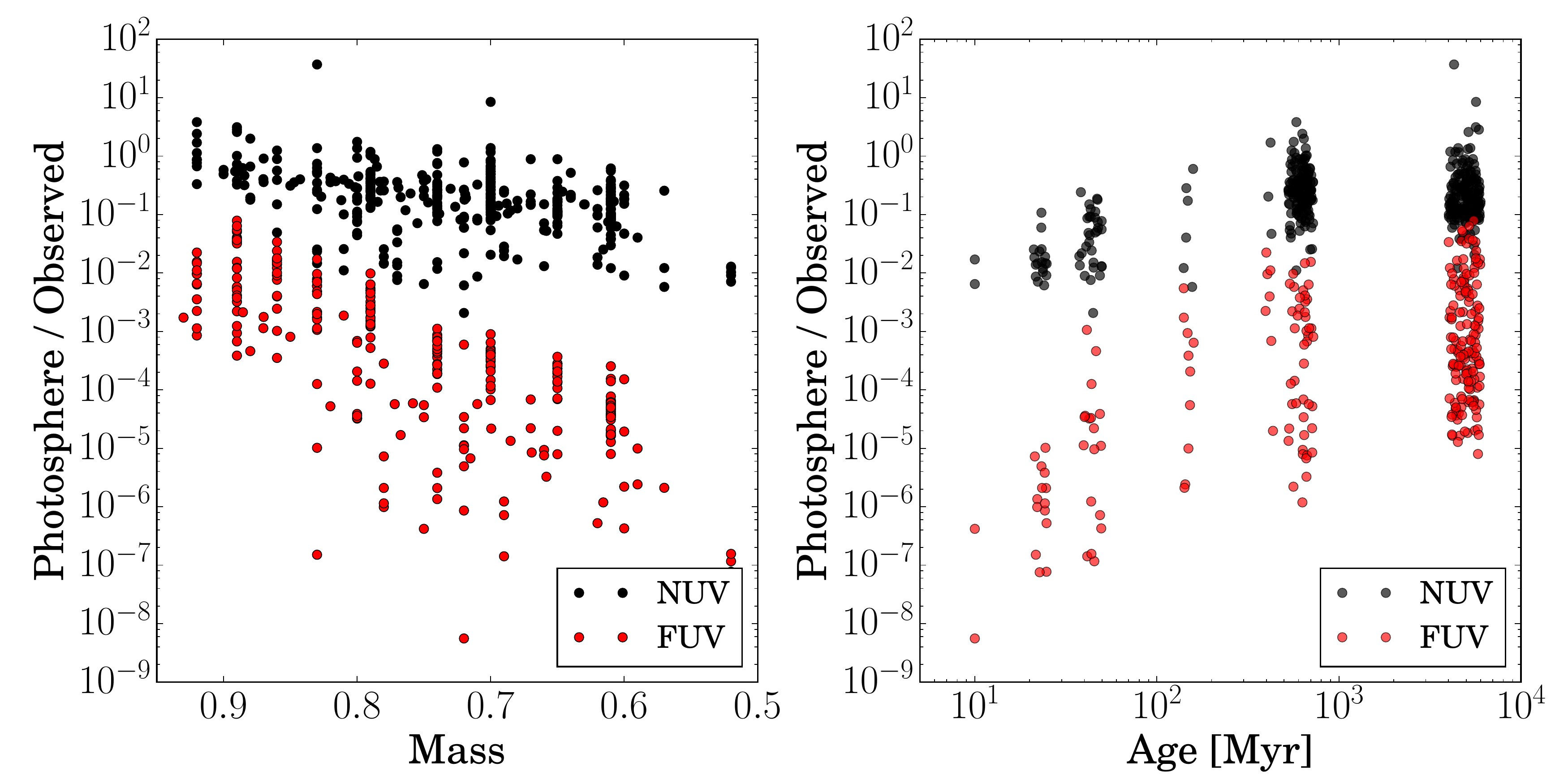}
\caption{Fraction of the photospheric NUV and FUV flux densities compared to the observed as a function of mass (left) and age (right). Fractions above 1 are caused by uncertainties in the stellar mass, age, or model and were taken to be equal to 1 in our analysis. Ages have been moved from abscissa for clarity. \label{fig:phot_obs}}
\vspace{1pt}
\end{figure*}

\section{The K Star Sample}\label{sec:sample}

\begin{figure*}[t]
\centering
\includegraphics[width=0.95\linewidth]{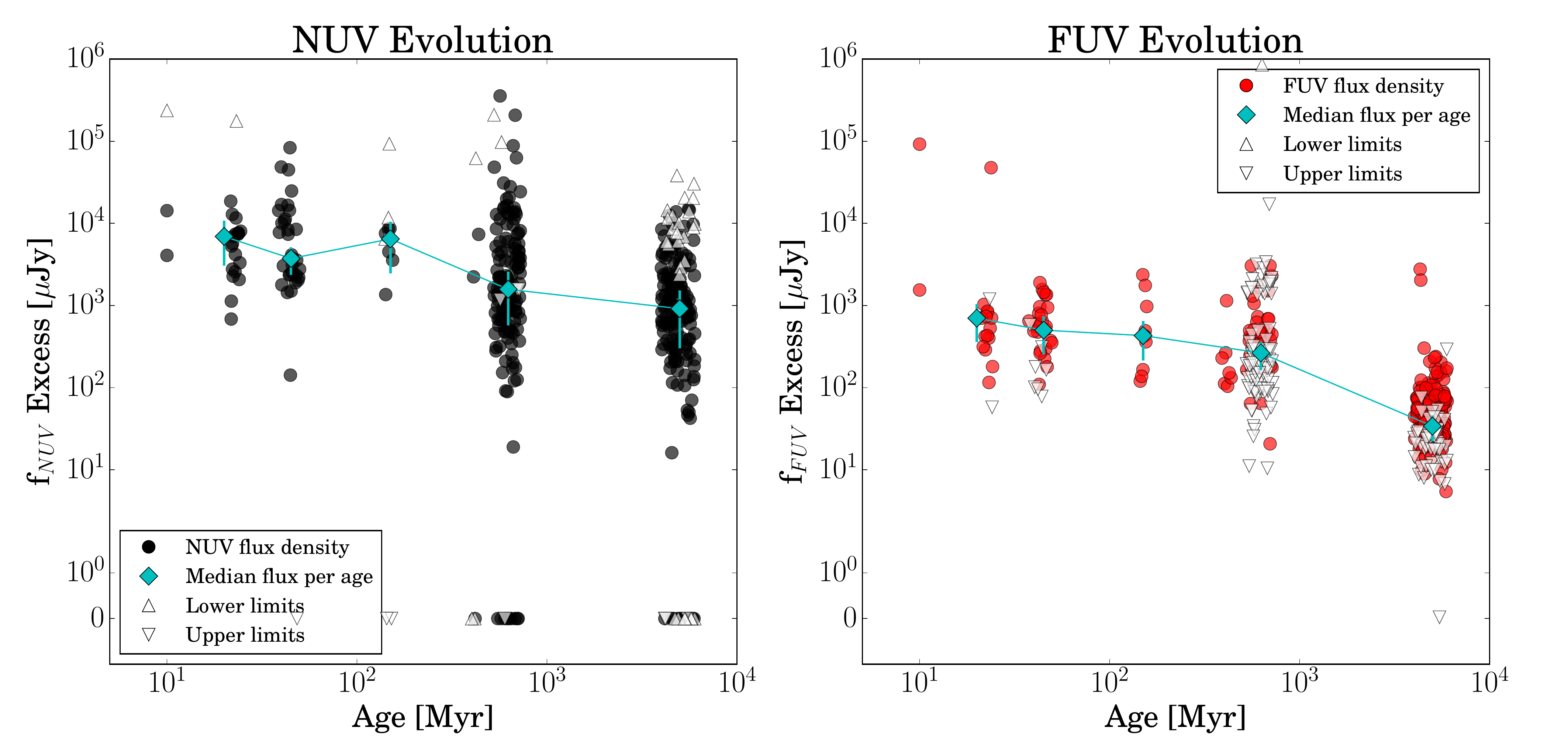}
\caption{Absolute NUV (left) and FUV (right) excess flux density (i.e. photosphere-subtracted). The cyan diamonds represent the median value of the detections, upper limits, and lower limits for each age group. The error bars represent inner quartiles. Upper limits are shown by downward triangles and lower limits with upward triangles. Ages have been moved from abscissa for clarity. Typical errors are smaller than the markers.  \label{fig:ffdensity_age}}
\vspace{10pt}
\end{figure*}

We compiled K stars in young moving groups (YMG) and clusters ranging in age from 10 Myr to 625 Myr. The associations included are TW Hydra (10 $\pm$ 3 Myr, \citealt{Bell2015}), Beta Pictoris (24 $\pm$ 3 Myr, \citealt{Shkolnik2017}), Columba (42$^{+6}_{-4}$ Myr, \citealt{Bell2015}), Carina (45$^{+11}_{-7}$ Myr, \citealt{Bell2015}), Tucana-Horologium (45 $\pm$ 4 Myr, \citealt{Bell2015}), AB Doradus (149$^{+51}_{-19}$ Myr, \citealt{Bell2015}), Ursa Majoris (414 $\pm$ 23 Myr;  \citealt{Jones15b, Jones17}), Praesepe (600 Myr, \citealt{Kraus2007}), and Hyades (625 $\pm$ 50 Myr, \citealt{Perryman1997}). In addition, we define a field star sample as K stars within 30 pc to which we assign an age of 5 Gyr. The sample was drawn from SIMBAD and confirmed individually to not be close binaries or known association members. 

Taking into account the wide range of ages in this study, we must consider how similar spectral types represent different masses over various ages. Using the effective stellar temperatures in \citet{Pecaut2013} and the model isochrones from \citet{Baraffe2015}, we determine mass estimates for the spectral types used in this work following Table \ref{tab:massest}. We used masses from 0.6 -- 0.9 M$_{\odot}$ but allowed one spectral subtype lower and higher to account for spectral type uncertainty.

The YMG members were identified by \citet{GagneTWA} for TW Hydra; \citet{Shkolnik2017} for Beta Pictoris; \citet{Malo2013} for Carina, Columba, and AB Doradus;  \citet{KrausTucHor} for Tucana Horologium; \citet{Montes2001} for Ursa Major; \citet{wang1995} for Praesepe; and \citet{goldmanhyades} for the Hyades. Due to the low number of K star members in some of the age ranges, we combined groups of similar age ranges to calculate the median and inner quartiles for each age bin, as seen in Figures \ref{fig:massdistributions_nuv} and \ref{fig:massdistributions_fuv}.

\section{GALEX Photometry}\label{sec:photometry}

We cross referenced proper motion corrected coordinates of our sample to \textit{GALEX} using the \textit{GALEX}view tool\footnote{http://galex.stsci.edu/galexview/} with a search radius of 10\arcsec. The NUV (1771 -- 2831 \AA) and FUV (1344 -- 1786 \AA) detectors have a non-linear response at 104 counts s$^{-1}$ and 34 counts s$^{-1}$, respectively, equivalent to a magnitude of $\approx$15 for both NUV and FUV. Measured magnitudes less than 15 were thus taken as lower limits. Additionally, we excluded detections with photometric flags for bright star window reflection, dichroic reflection, detector run proximity, or bright star ghost. We visually inspected the \textit{GALEX} tiles for each observation to ensure that there was no contamination that was not caught by the flags. A comparison of the number of objects that were in our input sample, observed by \textit{GALEX}, and were then resolved with no photometric flags for each band can be seen in Table \ref{tab:galex} and Figure \ref{fig:mass_fractions}.

In the NUV, the higher mass objects were less likely to have usable photometry, as they were too bright for reliable photometric measurements. This is due to the more massive stars being intrinsically brighter and saturating the detector. For the Hyades and Praesepe clusters, the low fraction of detections is most likely due to their distances (∼47 pc and 177 pc, respectively; \citealt{vanleeuwen}).

For some objects there were multiple observations with multiple exposure times, in which case we took the weighted mean of the magnitude as our measurement and the weighted standard deviations as our uncertainty. All of the photometry can be seen in Table \ref{tab:data}.

For the objects that were observed but not detected, we calculated an upper limit for the object by repeating the search with a 10\arcmin\space radius and proceeding using the method described in \citet{Schneider2018}. The objects that were measured with a magnitude $<$15 were taken as upper limits. These were then addressed in the same manner as the detections.

\begin{figure}[t]
\centering
\includegraphics[width=\linewidth]{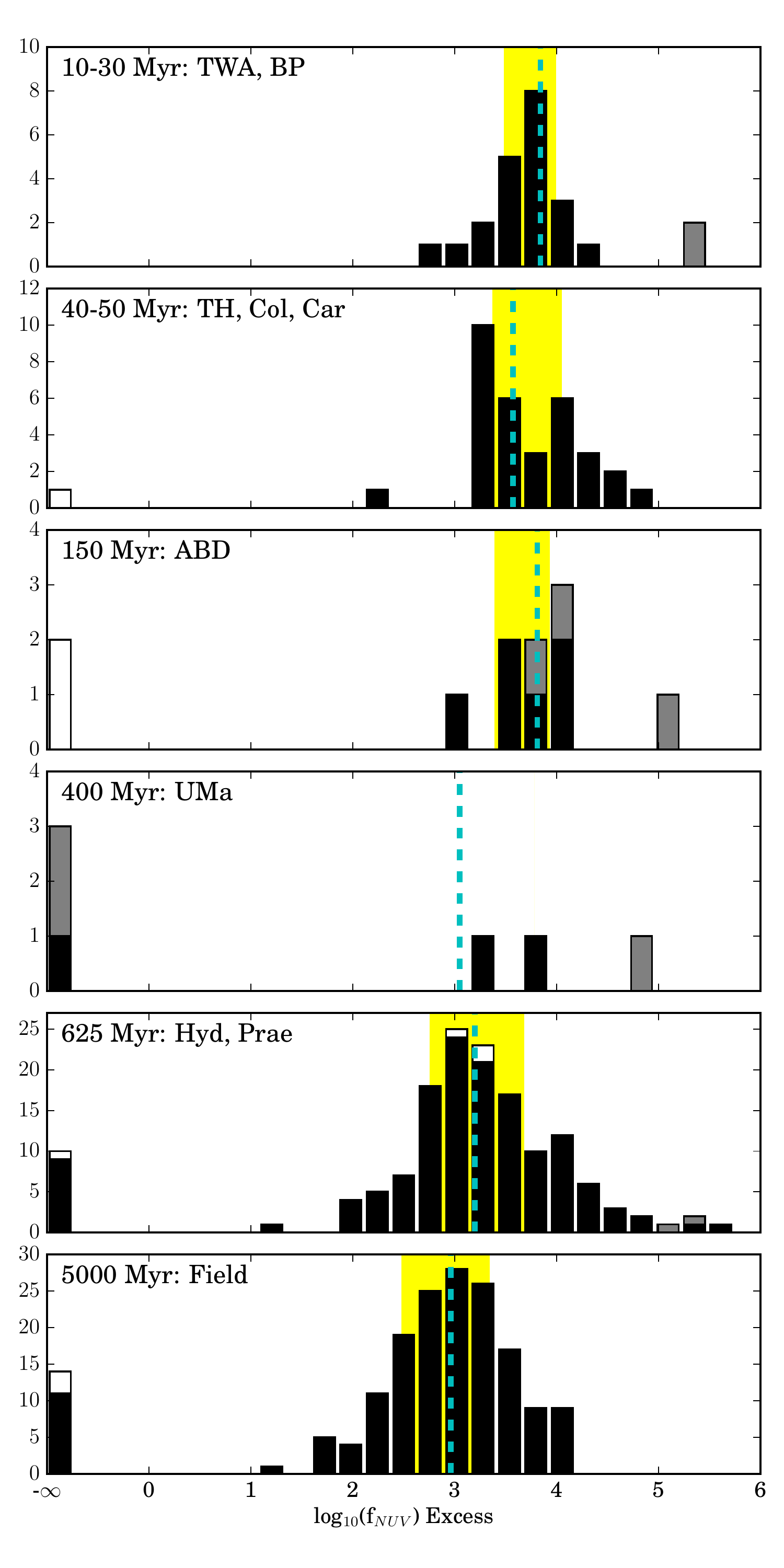}
\caption{Histogram of NUV excess flux densities. Median values are shown by cyan dashed lines. The yellow areas represent the inner quartiles. Upper limits are shown in white and lower limits in gray. The negative infinity values are stars with zero excess. \label{fig:histfd_nuv}}
\end{figure}

\begin{figure}[t]
\centering
\includegraphics[width=\linewidth]{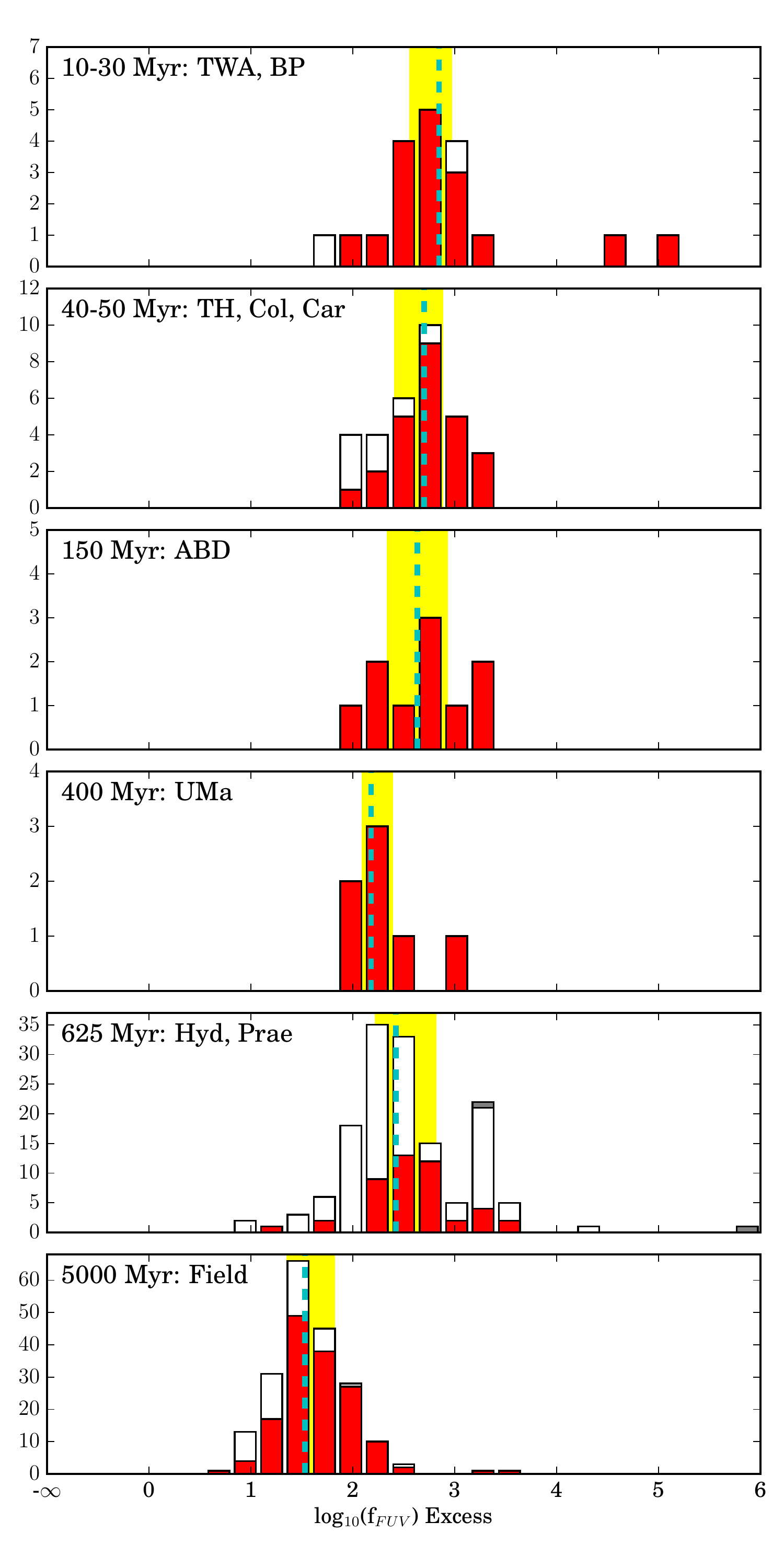}
\caption{Same as Figure \ref{fig:histfd_nuv} but for FUV excess flux density. Median values are shown by cyan dashed lines. The yellow areas represent the inner quartiles. Upper limits are shown in white and lower limits in gray. \label{fig:histfd_fuv}}
\end{figure}

\begin{figure*}[t]
\centering
\includegraphics[width=0.95\linewidth]{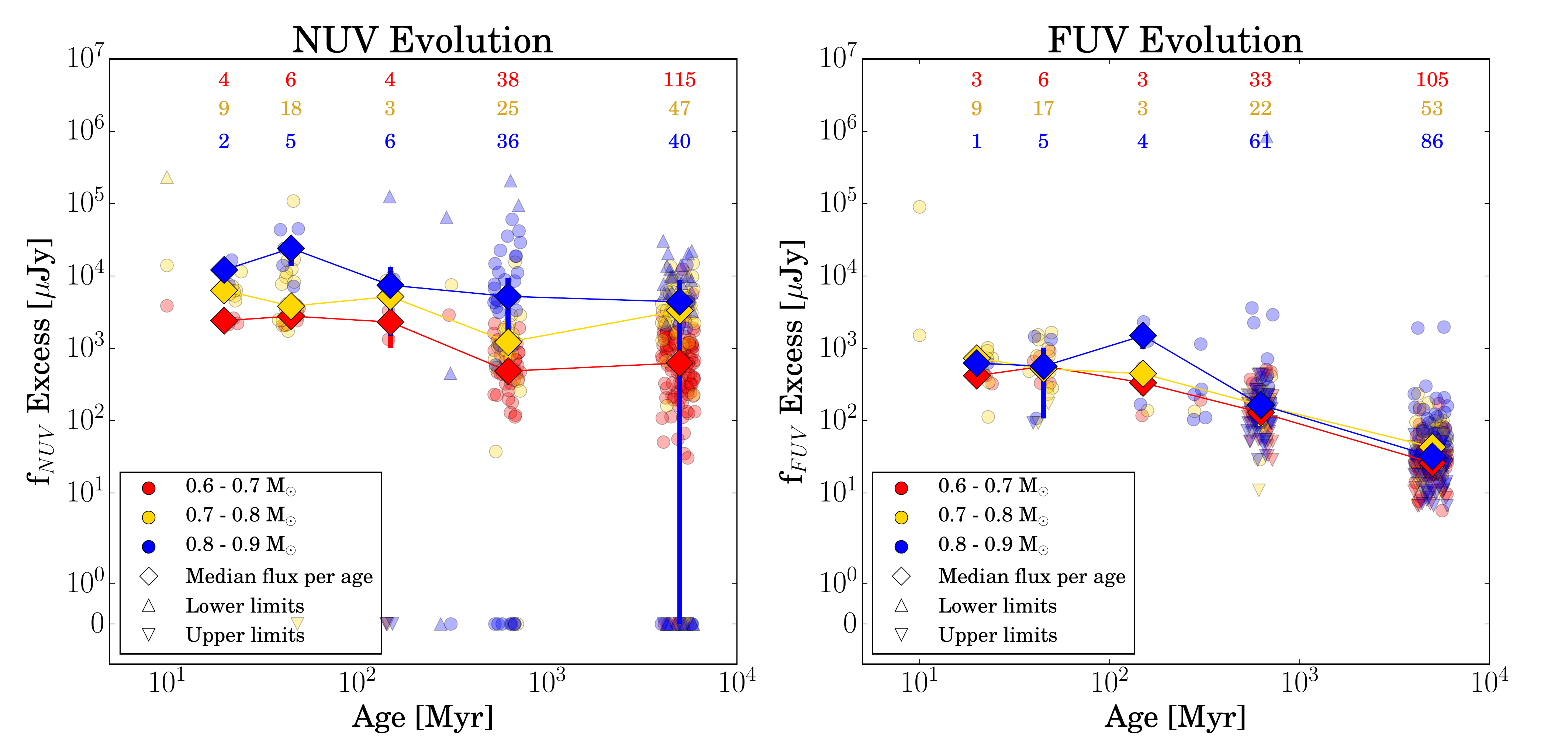}
\caption{Absolute NUV (left) and FUV (right) excess flux density (i.e. photosphere-subtracted) separated by mass. The diamonds represent median values and the lines the inner quartiles. The large inner quartile for the early K stars at the field age is due to a large number of stars without excess flux. The number of stars in each age and mass bin are shown at the top. Typical errors are smaller than the markers. \label{fig:masses_review}}
\end{figure*}

\begin{figure}[t]
\centering
\includegraphics[width=\linewidth]{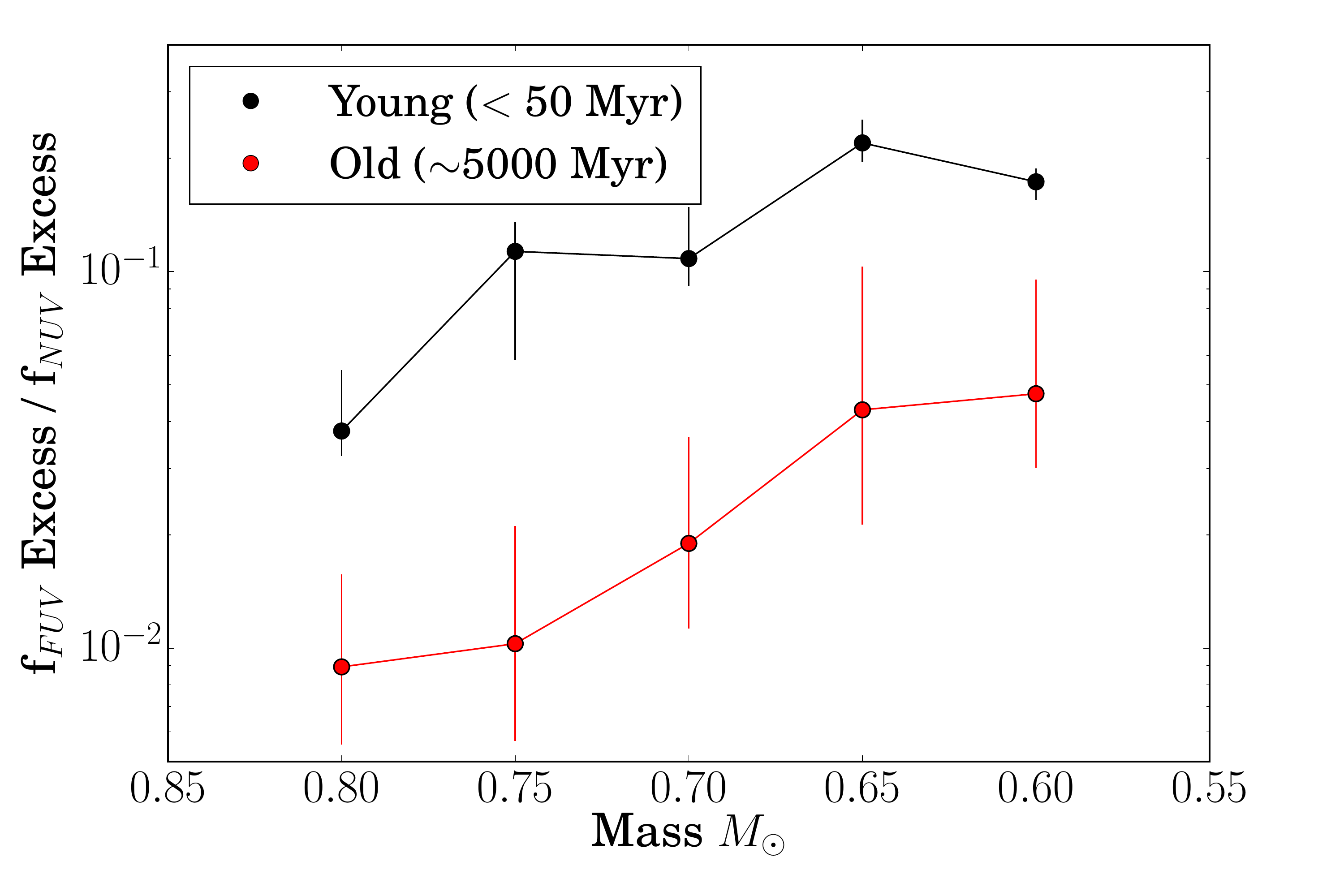}
\caption{Ratio of FUV Excess to NUV Excess as a function of mass. The ratio is clearly mass and age dependent. \label{fig:fuv_nuv_vs_mass}}
\vspace{10pt}
\end{figure}

\begin{figure}[t]
\centering
\includegraphics[width=\linewidth]{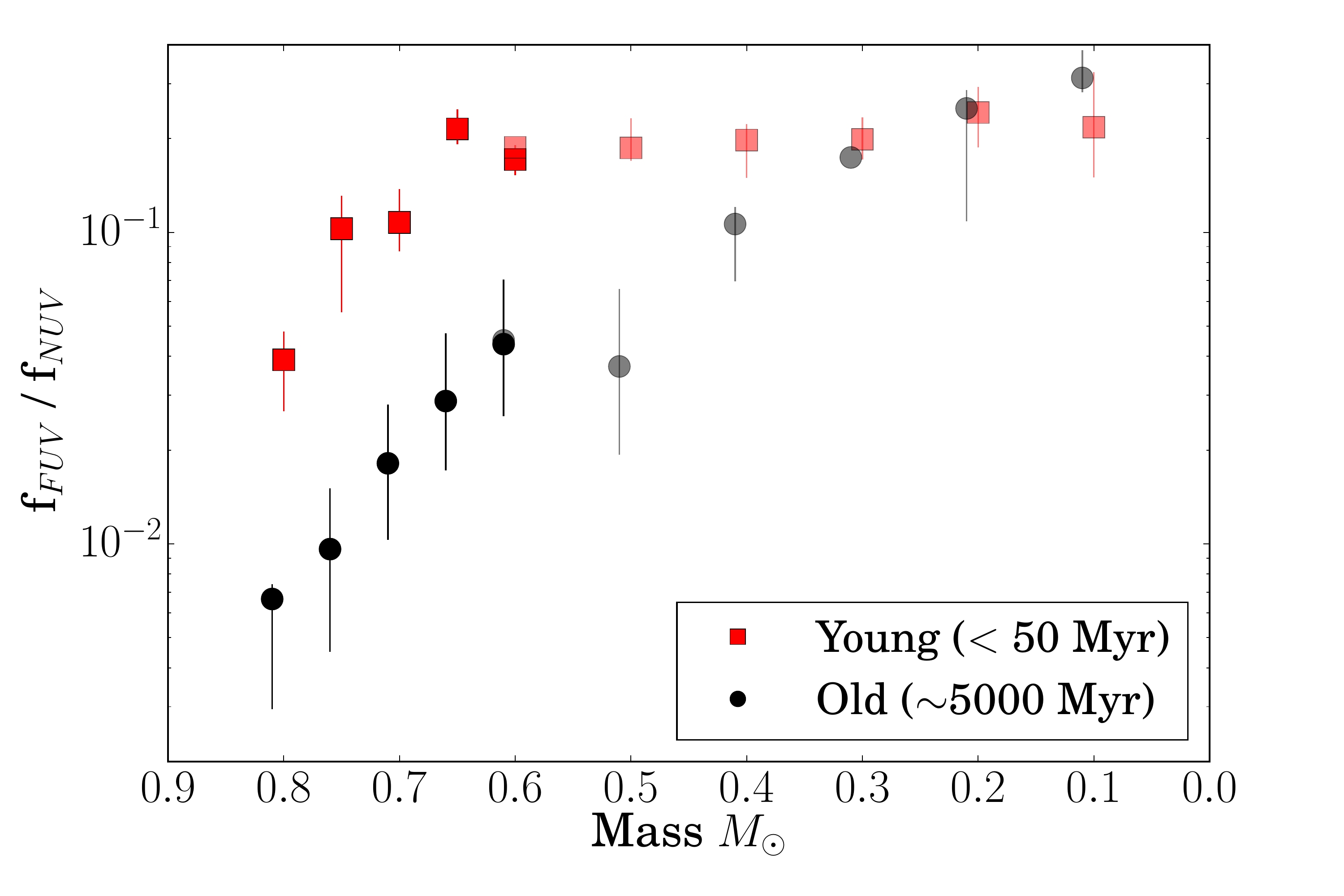}
\caption{Ratio of FUV to NUV as a function of mass for K stars (opaque) from this work and M stars (translucent) from \citet{Schneider2018}. In this case the photosphere has not been subtracted for either K or M stars.  \label{fig:fuv_nuv_vs_mass_no_excess}}
\end{figure}

\section{Evolution of the Photospheric UV Emission}\label{sec:photoevolution}

One of the goals of the HAZMAT program is to provide measurements of FUV and NUV flux densities for low-mass stellar models. Most low-mass stellar atmosphere models predict photospheric emission only and do not contain contributions from the stellar upper-atmosphere, leading to an underestimate of stellar photospheric emission. Work is being carried out to include contributions from these regions (i.e. \citealt{peacock}, \citealt{fontenla}), but observations are needed to fully inform these models with empirical data in the NUV and FUV.

The PHOENIX stellar atmosphere models (\citealt{hauschildt1997}, \citealt{short2005}) were used to calculate the photospheric NUV and FUV flux densities of each K star in our sample using the stellar masses derived in Table \ref{tab:massest} and the age of the star. The evolution of the NUV and FUV photospheric model flux densities for select masses from 0.6 to 0.9 M$_{\odot}$ with a reference 1 M$_{\odot}$ can be seen in Figure \ref{fig:mvd_vs_age}. As the mass increases, the photospheric contribution becomes greater. For each mass, the photospheric contributions become nearly constant after the age at which contraction onto the main sequence has ended, since the temperature of the star does not significantly change after the pre-main sequence evolution during the ages used in this study (Figure \ref{fig:hr}).

Figure \ref{fig:phot_obs} shows the fraction of the photospheric flux density compared to the absolute observed values for the YMG members and field stars as a function of both mass and age. The NUV has a higher fraction of photosphere compared to observed flux, with most of the contribution being between 10 and 100\%. Values reported that were larger than 100\% are due to uncertainties in both the mass and age of the star and were taken to be 100\% photospheric. The photospheric FUV flux density was typically $<$1\%. For both the NUV and the FUV we see a decrease in photospheric flux with decreasing mass, although this trend is much steeper for the FUV, where the photospheric contribution becomes negligible. With increasing age, we see an increase in photospheric contribution equally for both NUV and FUV flux densities. 

For each of the YMG members and field stars, we subtract the photospheric contribution derived from the PHOENIX models from the absolute observed \textit{GALEX} flux densities to calculate the excess emission, representative of upper-atmosphere activity. 

\section{Evolution of the Observed NUV and FUV Emission}\label{sec:uv evolution}

With the second data release of \textit{Gaia} \citep{gaiadr2},  accurate distances to all objects in our sample are available. Therefore, we investigate the absolute \textit{GALEX} NUV and FUV flux densities rather than analyzing them relative to the Two Micron All-Sky Survey (2MASS) J-band, such as in \citet{Shkolnik2014} and \citet{Schneider2018}.

To explore the evolutionary trends, we convert the \textit{GALEX} reported magnitudes to flux densities in $\mu$Jy using

\begin{equation}
    f_{GALEX} = 10^{\frac{23.9 - m_{GALEX}}{2.5}}
\end{equation}

where $m_{GALEX}$ is the \textit{GALEX} FUV or NUV magnitude. These values were then translated to absolute fluxes using the known distances from \textit{Gaia} Data Release 2 \citep{gaiadr2}. Because we are interested in the excess UV contribution from the K stars, we calculated the photospheric contribution using the PHOENIX models as described in section \ref{sec:photoevolution} and subtracted these from the flux density calculations reported in \textit{GALEX}. 

Figure \ref{fig:ffdensity_age} shows the evolution of the excess NUV and FUV flux densities as a function of age. Both the NUV and FUV decrease in time. The slope of the decrease is much more gradual for the NUV than that of the FUV and the values obtained for early M stars in \citet{Shkolnik2014} and mid- to late M stars in \citet{Schneider2018}. Unlike both \citet{Shkolnik2014} and \citet{Schneider2018} where the median values remain constant up until 625 Myr and then distinctly drop off, we see a more gradual decrease in time starting at 100 Myr in the FUV. The wide spread comes from intrinsic astrophysical variation and perhaps initial spin rate, which has been shown to cause large spreads in the X-ray due to the bimodal distribution of stellar rotation in young open clusters for solar-like stars \citep{gondoin17, gondoin18}. Additionally, the uncertainty in stellar ages beyond the age of the Hyades may cause increased variation among the field stars.

Figures \ref{fig:histfd_nuv} and \ref{fig:histfd_fuv} show the distribution of the NUV and FUV excess flux for the different age groups. The scatter in the FUV ranges from 1.5 -- 3 orders of magnitude, similar to \citet{Shkolnik2014}. \citet{Shkolnik2014} and \citet{Schneider2018} see a change in the median FUV value of about 1.5 orders of magnitude for stars with masses 0.35 -- 0.6 M$_{\odot}$, whereas for the K stars we see a shift of only 1 order of magnitude. For the NUV, the scatter ranges from 1 -- 4 orders of magnitude, with the scatter increasing for the older age groups. The slopes of the NUV and FUV excess flux density vs age curves are found in Table \ref{tab:coeff}. These trends can be used to predict flux densities and their ranges for K stars in the NUV and FUV bands. 

\newpage
\subsection{Mass Distributions}

Figure \ref{fig:masses_review} is the same as Figure \ref{fig:ffdensity_age} but broken down by mass group. In both the NUV and FUV, we see that in general, all three mass groups follow the trends of each other. This is counter to the trends seen in \citet{Schneider2018}, where the late M stars remain active much longer than early M stars. Even though the higher-mass stars have a larger contribution from the photosphere that is being subtracted off to yield the excess, the intrinsic flux of the higher-mass stars is large enough that the excess flux is still largest for higher-mass stars throughout most of the evolutionary period. However, once the stars reach field age, the photospheric contribution becomes so large that most of the excess flux for higher-mass stars approaches zero. 

\begin{figure*}[t]
\centering
\includegraphics[width=\linewidth]{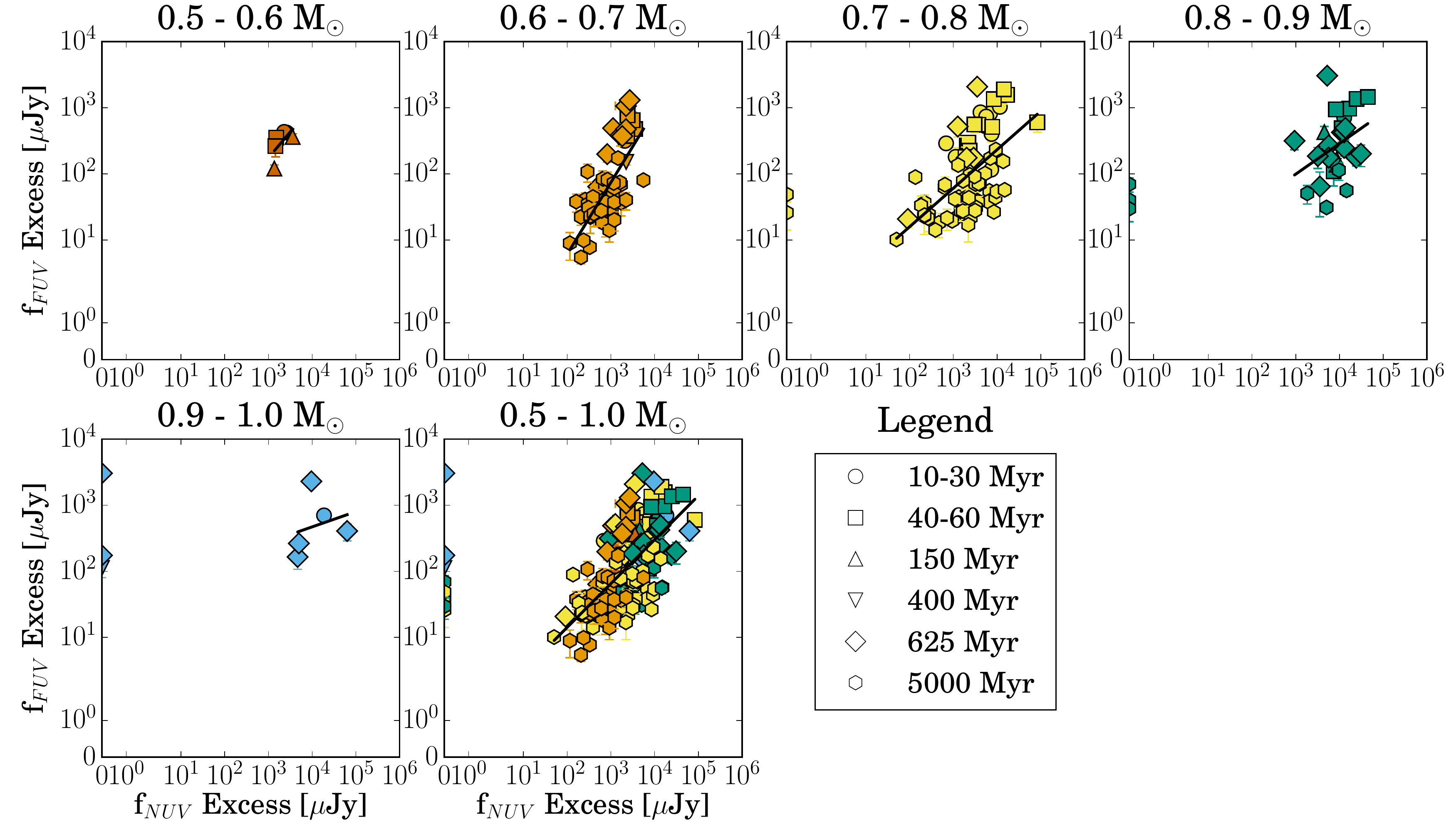}
\caption{FUV excess flux density versus NUV excess flux density for different mass bins. The parameters for the best fit lines can be found in Table \ref{tab:coeff}. The zero excess points were not included in the calculation of the trend lines. Typical errors are smaller than the markers. \label{fig:nuv_vs_fuv}}
\vspace{10pt}
\end{figure*}

\section{The Relationship Between \textit{GALEX} FUV and NUV for K Stars}\label{sec:fuv nuv relationship}

\subsection{FUV $/$ NUV}\label{sec:fuv/nuv}

The ratio of FUV to NUV is valuable in understanding the photochemistry of terrestrial planet atmospheres. \citet{Segura2005} showed that lifetimes of the biogenic gases CH$_4$, N$_2$O, and CH$_3$Cl are actually increased, allowing a higher probability of being detected. However, the FUV to NUV ratio also sets the abundance rates of molecules such as CO$_2$, O$_2$, and O$_3$, as the FUV will break down CO$_2$, forming O$_2$ and O$_3$, the latter of which will be destroyed by the NUV. This abiotic oxygen and ozone can thus produce false-positive biosignatures \citep[e.g.][]{DG14, tian14, harman15}. 

We calculate FUV/NUV excess flux densities as a function of mass for stars with both NUV and FUV flux densities, as seen in Figure \ref{fig:fuv_nuv_vs_mass}. The trends seen in \citet{Schneider2018} continue into the K-star regime, with increasing separation between the young and old stars as the stellar mass increases, demonstrating that the FUV/NUV ratio is both mass- and age-dependent. 

\subsection{NUV vs FUV}\label{sec:nuv vs fuv}

\begin{figure}[t]
\centering
\includegraphics[width=\linewidth]{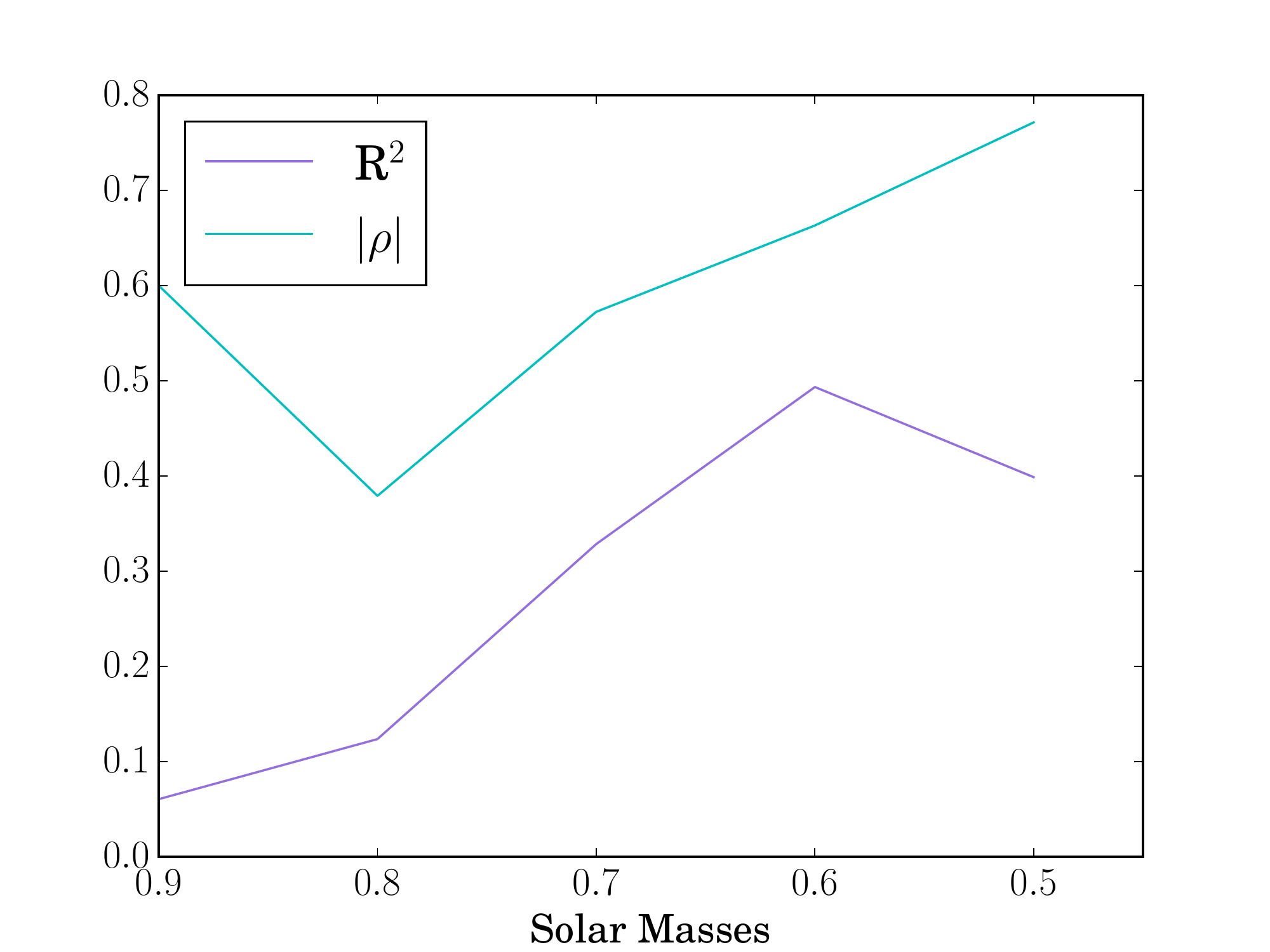}
\caption{Pearson's R$^2$ correlation statistic and Spearman rank correlation for the trends seen in Figure \ref{fig:nuv_vs_fuv}. The correlations become worse at higher mass most likely due to the larger variation of photospheric contribution in age, as seen in Figure \ref{fig:mvd_vs_age}.   \label{fig:r2_vs_spt}}
\end{figure}

When comparing the NUV to FUV excess flux density, we do not see the same tight correlations as \citet{Shkolnik2014} nor \citet{Schneider2018}. Instead, we see in Figure \ref{fig:nuv_vs_fuv} that with increasing mass (i.e. earlier spectral type), the scatter becomes greater and the correlation worsens. We calculated both a Pearson’s R$^2$ statistic \citep{pearsonr2} and a Spearman Rank correlation value \citep{Spearman1904} for each K spectral type mass range. The results can be seen in Figure \ref{fig:r2_vs_spt}. In both tests we see a similar trend, where the highest mass (i.e. earliest spectral type) stars have correlations generally with $\rho$ $<0.5$ and lower mass (i.e. later spectral type) stars have correlations generally $\rho$ $>0.5$. This trend also appears in \citet{Miles2017}.

The reason for this decrease in correlation with higher mass can most easily be seen in analyzing the model photospheric fluxes in Figure \ref{fig:mvd_vs_age}. In comparing the model values of a single mass from 10 Myr -- 5 Gyr, the photospheric flux density of a 0.6 M$_{\odot}$ star changes by no more than a single order of magnitude. However, the photospheric flux density of a 0.9 M$_{\odot}$ star is much more prevalent and changes by as much as four orders of magnitude. The scatter in the FUV excess flux density vs NUV excess flux density then comes from differences in age, whereas the intrinsic scatter of a single age group is much smaller.

\begin{figure}[th]
\centering
\includegraphics[width=\linewidth]{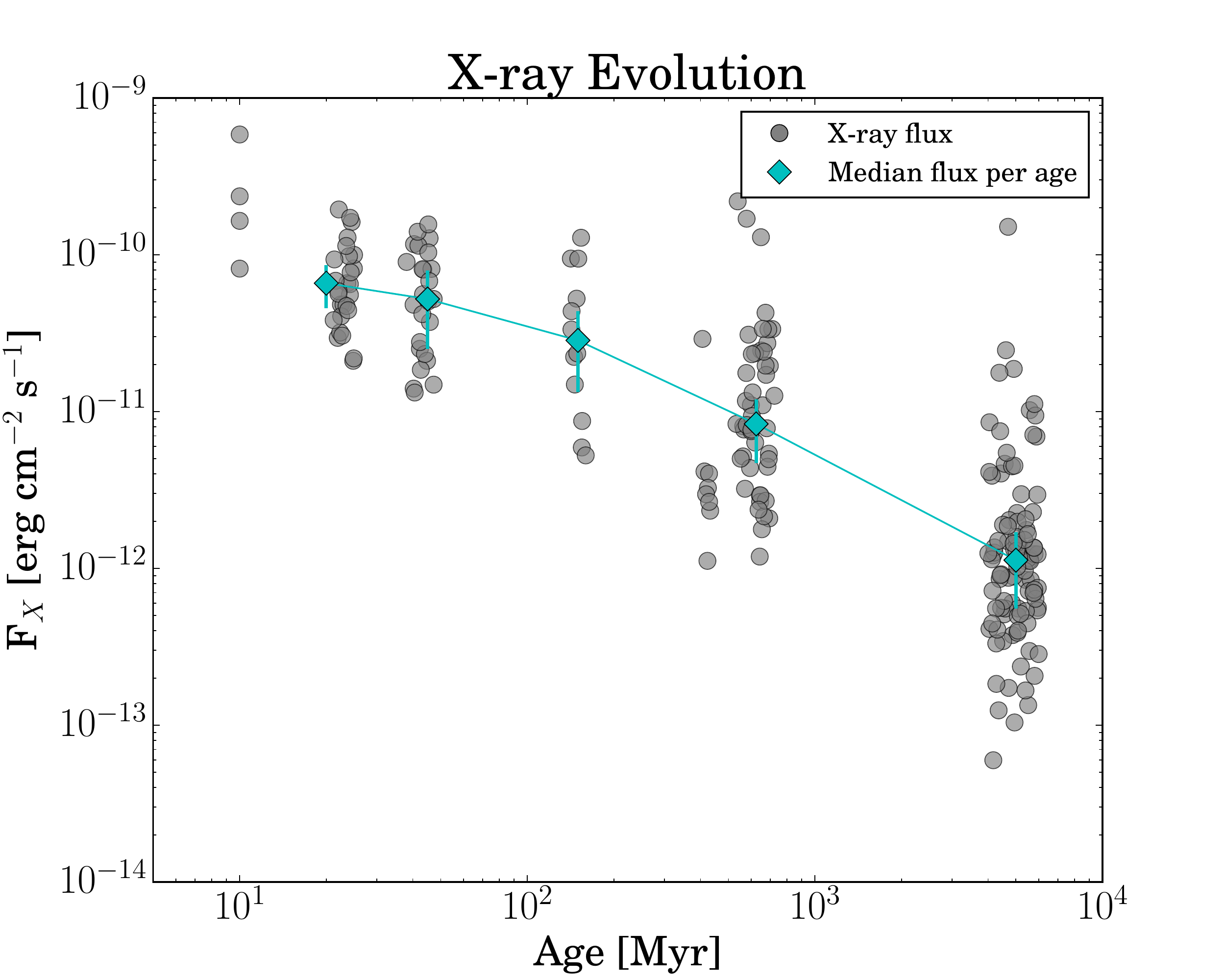}
\caption{Same as Figure \ref{fig:ffdensity_age} but for X-ray flux (as compared to flux density). \label{fig:xray_evolution}}
\end{figure}

\begin{figure}[t]
\centering
\includegraphics[height=0.9\textwidth]{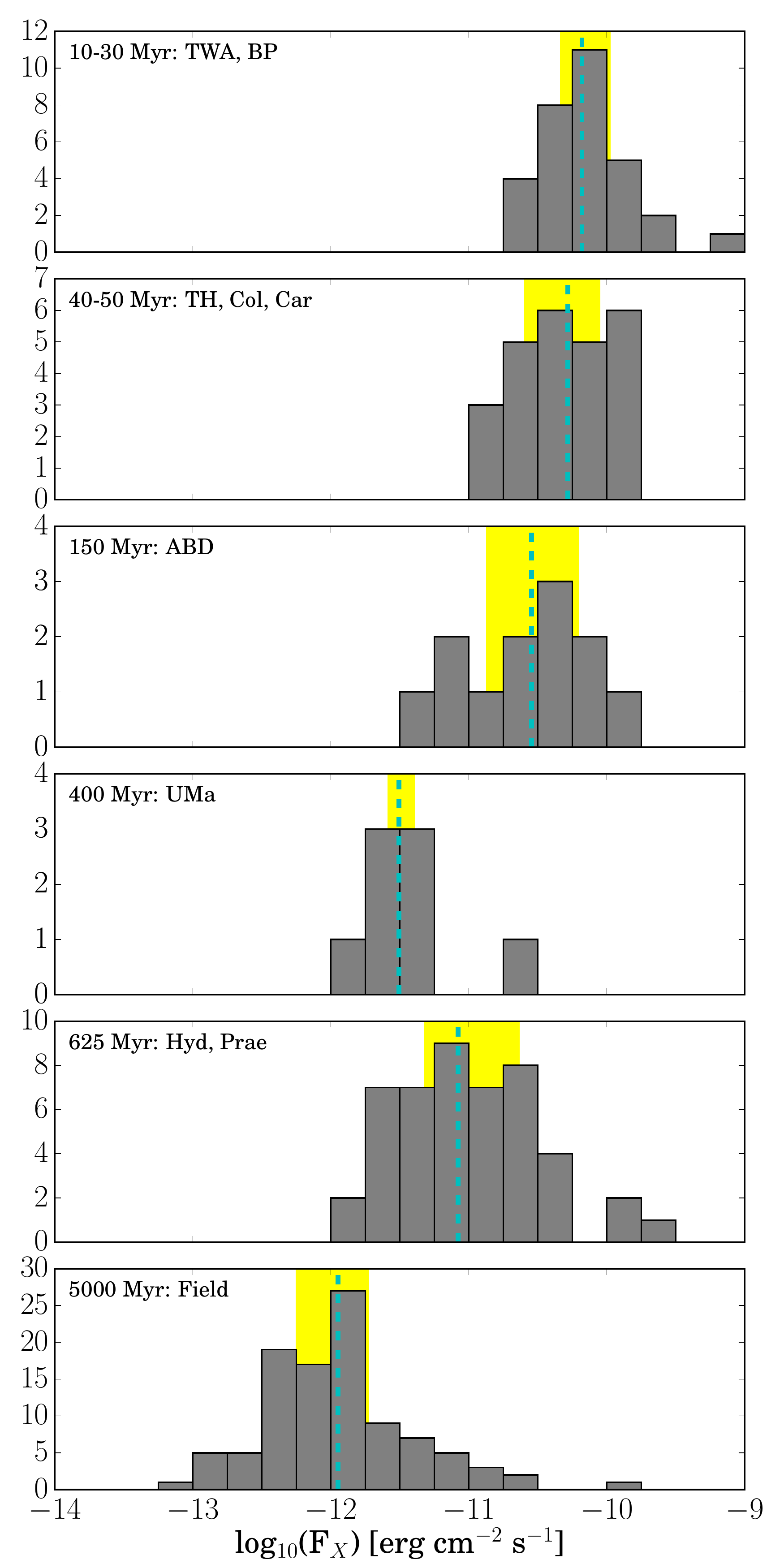}
\caption{Same as Figures \ref{fig:histfd_nuv} and \ref{fig:histfd_fuv} but for X-ray flux. Median values are shown by cyan dashed lines. The yellow areas represent the inner quartiles. \label{fig:histfd_xray}}
\end{figure}

\begin{figure}[th]
% \vspace{-1cm}
\centering
\includegraphics[width=\linewidth]{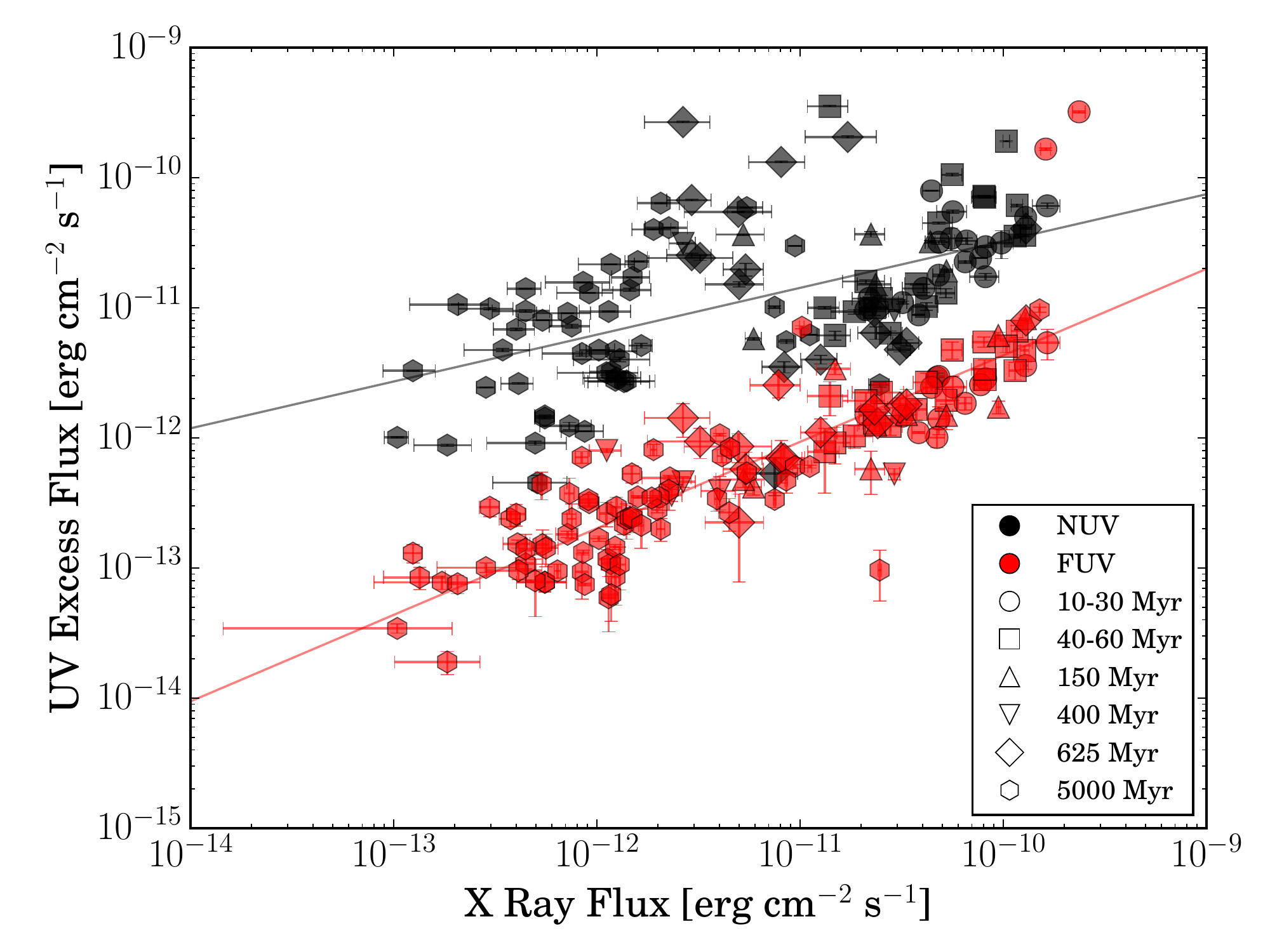}
\caption{NUV and FUV Excess flux against X-ray flux from \textit{ROSAT} for those stars with detections in both. Equations for the two fits are reported in Table \ref{tab:coeff}. \label{fig:uv_vs_xray}}
\end{figure}

\begin{deluxetable*}{c c c c c c c}[t]
\tabletypesize{\footnotesize}
\tablecaption{\normalsize{Fits to $\rm{log}(y) = m*log(x) + b$ in logarithmic space.} 
\label{tab:coeff}}
\tablecolumns{8}

\tablehead{
\colhead{log(x)} & \colhead{log(y)} & \colhead{Subset} & \colhead{Slope (m)}   & \colhead{Intercept (b)}   & \colhead{R$^2$} & \colhead{No.\tablenotemark{a}}\\
\colhead{[$\mu$Jy or Myr]} & \colhead{[$\mu$Jy]} & \colhead{} & \colhead{}   & \colhead{}   & \colhead{} & \colhead{}
}
\startdata
NUV Excess & FUV Excess & 0.5 - 0.6 M$_{\odot}$ & 0.784 $\pm$ 0.177 & -0.107 $\pm$ 0.594 & 0.399 & 6 \\
NUV Excess & FUV Excess & 0.6 - 0.7 M$_{\odot}$ & 1.086 $\pm$ 0.039 & -1.387 $\pm$ 0.120 & 0.493 & 61 \\
NUV Excess & FUV Excess & 0.7 - 0.8 M$_{\odot}$ & 0.583 $\pm$ 0.025 & 0.033 $\pm$ 0.091 & 0.328 & 75 \\
NUV Excess & FUV Excess & 0.8 - 0.9 M$_{\odot}$ & 0.463 $\pm$ 0.074 & 0.605 $\pm$ 0.297 & 0.124 & 27 \\
NUV Excess & FUV Excess & 0.9 - 1 M$_{\odot}$ & 0.231 $\pm$ 0.123 & 1.750 $\pm$ 0.502 & 0.061 & 5 \\
NUV Excess & FUV Excess & All & 0.663 $\pm$ 0.016 & -0.174 $\pm$ 0.056 & 0.418 & 174 \\
X-ray & NUV Excess & All & 0.360 $\pm$ 0.009 & -6.888 $\pm$ 0.097 & 0.281 & 110 \\
X-ray & FUV Excess & All & 0.665 $\pm$ 0.012 & -4.717 $\pm$ 0.130 & 0.745 & 130 \\
Age & NUV Excess & 10 Myr - 150 Myr & 0.003 $\pm$ 0.472 & -10.635 $\pm$ 0.836 & 0.000 & 3 \\
Age & NUV Excess & 150 Myr - 5 Gyr & -0.536 $\pm$ 0.343 & -9.498 $\pm$ 0.976 & 0.877 & 3 \\
Age & FUV Excess & 10 Myr - 625 Myr & -0.266 $\pm$ 0.257 & -11.280 $\pm$ 0.495 & 0.968 & 4 \\
Age & FUV Excess & 625 Myr - 5 Gyr & -0.987 $\pm$ 0.426 & -9.275 $\pm$ 1.414 & 1.000 & 2 \\
Age & X-ray & 10 Myr - 150 Myr & -0.424 $\pm$ 0.467 & -9.612 $\pm$ 0.746 & 0.981 & 3 \\
Age & X-ray & 150 Myr - 5 Gyr & -0.923 $\pm$ 0.285 & -8.522 $\pm$ 0.881 & 0.999 & 3 
\enddata
\tablenotetext{a}{Number of data points used in each fit.}
\end{deluxetable*}
\normalsize

\begin{figure*}[h]
\centering
\includegraphics[height=0.32\textheight, width=0.8\linewidth]{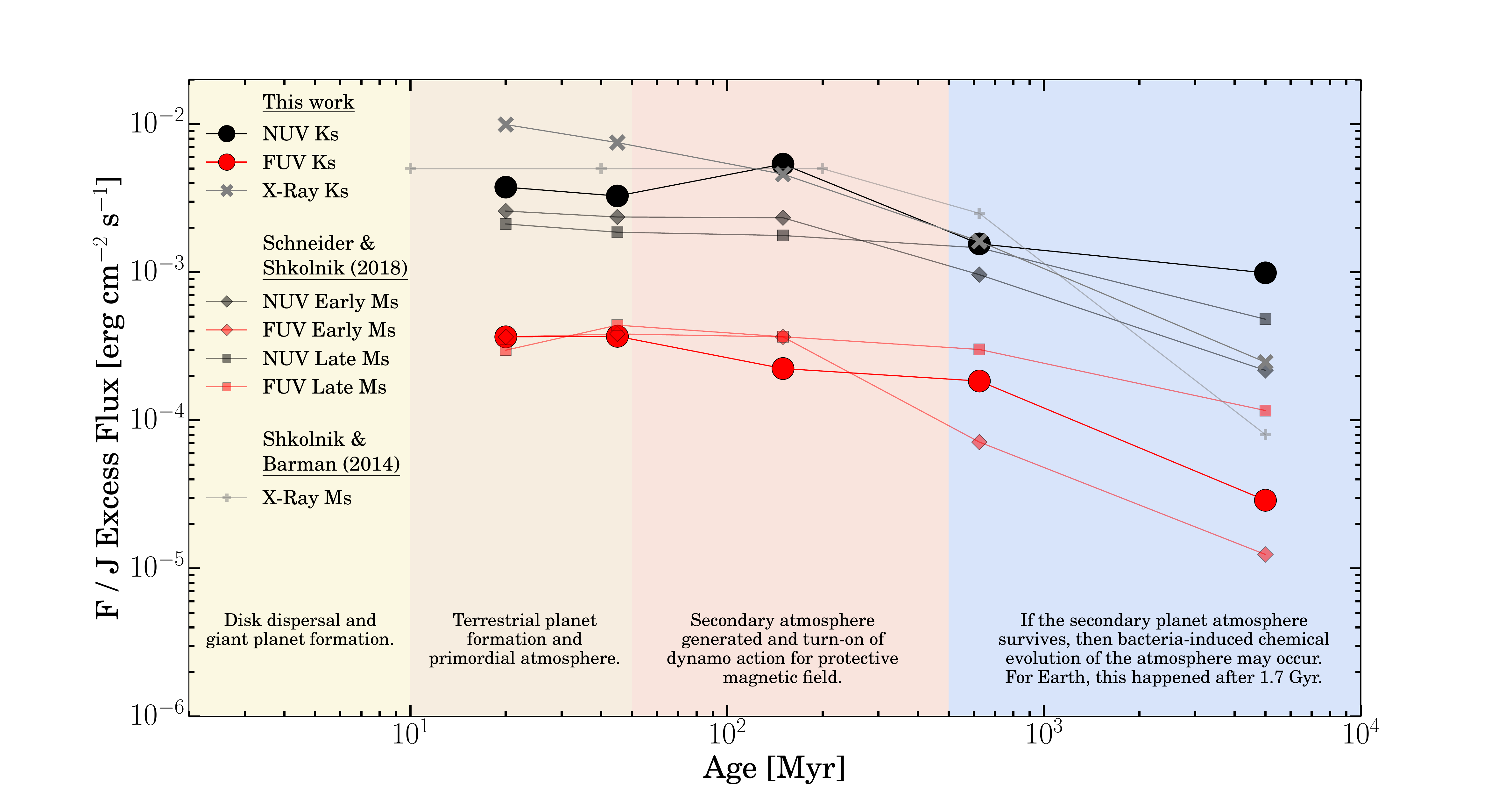}
\caption{Median NUV and FUV fluxes as a function of age comparing the K stars in this work to the M stars in \citet{Schneider2018}. The analysis of the K star NUV, FUV, and X-Ray evolution was redone similar to \citet{Shkolnik2014} and \citet{Schneider2018}, using the J band flux as a normalization for distance. The colored areas represents key evolutionary periods in a planet's formation. \label{fig:comp_planet_timeline}}
\end{figure*}

\begin{figure*}[h]
\centering
  \includegraphics[height=0.5\textheight, width=0.95\linewidth]{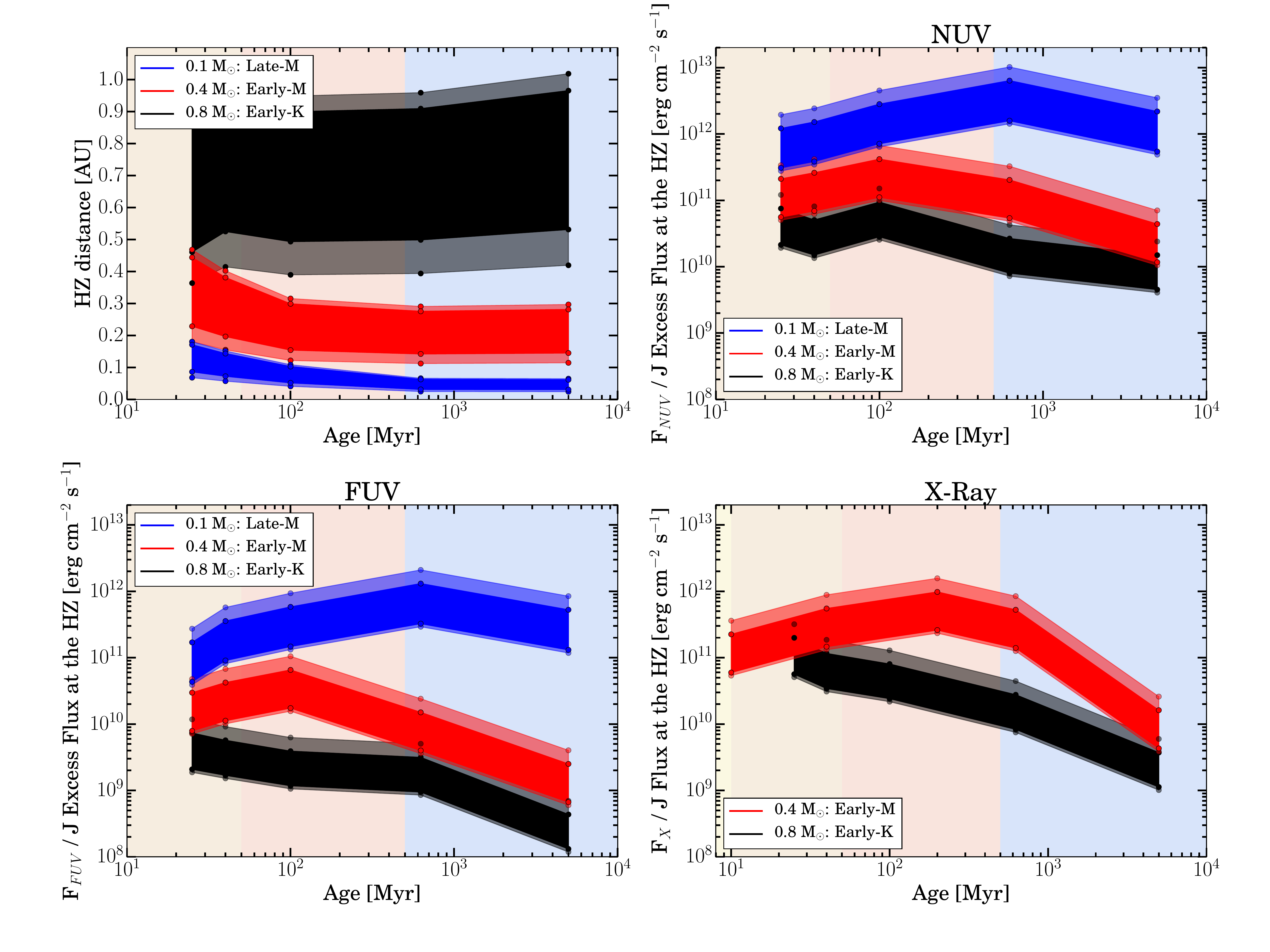}
  \caption{(a) Habitable zone distance from the star for a K star (0.8 M$_{\odot}$), an early-M star (0.4 M$_{\odot}$), and a late-M star (0.1 M$_{\odot}$). The opaque regions represent the conservative habitable zones and the translucent regions are the optimistic habitable zones. The color regions correspond to key planetary evolutionary periods outlined in Figure \ref{fig:comp_planet_timeline}. (b, c, d) NUV, FUV, and X-Ray median excess flux in the conservative (opaque) and optimistic (translucent) habitable zone ranges, respectively.}
  \label{fig:hz}
\end{figure*}

\section{X-Ray Evolution}\label{sec:xray evolution}

The X-ray flux has frequently been used as a stellar activity diagnosis \citep[e.g.][]{jackson12J, booth17}. X-ray wavelengths may be used in conjunction with the NUV and FUV to interpolate into the EUV, which is most powerful in understanding the photodissociation of molecules in a planet's atmosphere. To compare X-ray and UV data, we cross-referenced our sample of K stars with the Second ROSAT All-Sky Survey Point Source Catalog (2RXS, 5-124 \AA; \citealt{boller}) with a search radius of 38\arcsec, which was determined by \citet{voges1999} to be the 3$\sigma$ positional error. We then used the hardness ratio and the count rate to convert to flux $F_X$ in  in erg s$^{-1}$ cm$^{-2}$ using the empirical fit of \citet{schmitt1995}. The number of detections can be seen in the last column of Table \ref{tab:galex}. 

As can be seen in Figure \ref{fig:xray_evolution}, the X-ray flux remains steady through the age of Tucana-Horologium (45 Myr) and decreases by the age of AB Doradus (150 Myr), similar to the trend in FUV. These results are also seen in \citet{jackson12J}, who see that FGKM stars remain saturated in the X-ray until about 100 Myr. We derive a slope of $(-0.906 \pm 0.079)$log$(t)$ from 150 Myr - 5 Gyr (Table \ref{tab:coeff}). 

A histogram of the range in X-ray values for each age can be seen in Figure \ref{fig:histfd_xray}. The flux for each age ranges from 1-3 orders of magnitude, similar to the NUV and FUV values. Again, the wide spread likely comes from the bimodal distribution of stellar rotation in young open clusters for solar-like stars \citep{gondoin17, gondoin18} and uncertainty in stellar ages up to 10 Gyr among the field stars.

Unlike for the early-M stars in \citet{Shkolnik2014}, we do not see a clear distinction between young X-ray emitters and old emitters when looking at the UV flux compared to the X-ray flux  (Figure \ref{fig:uv_vs_xray}). This may be due to larger spread in x-ray flux  we see at the younger ages of K stars (Figure \ref{fig:histfd_xray}) and the more gradual decrease of the x-ray flux with age (Figure \ref{fig:comp_planet_timeline}) as compared to the M stars in  \citet{Shkolnik2014}, leading to an undefined distinction between young and old emitters. 

Figure \ref{fig:comp_planet_timeline} shows both the K and M star NUV, FUV, and X-ray fluxes (as opposed to flux densities) from this work, \citet{Shkolnik2014}, and \citet{Schneider2018}. The M-star work was done relative to the J band flux, as the distances to the stars were not all known. We have therefore redone this work relative to the J band flux for comparison with the M stars. The results are extremely similar except for the K dwarf NUV median value at 150 Myr, which is noticeably elevated. This is most likely due to small numbers in that age bin and uncertainties within the model photosphere, yet is still consistent within the inner quartiles seen in Figure \ref{fig:ffdensity_age}

Comparing the K stars to the M stars, we see in both the FUV and the X-ray that at the age where a planet is forming its atmosphere \citep[10 - 500 Myr; see][and references therein]{2010Icar..208..438S}, the K star flux has dropped below that of the M stars, meaning that any primordial atmosphere has a lower chance of being destroyed. In comparing the X-ray medians to those of the NUV and FUV, we see that the X-ray flux dominates during the young ages of the stars, then falls below the NUV values after 650 Myr, a trend also seen in the M dwarfs of \citet{Shkolnik2014}.

\section{UV and X-Ray Flux at the Habitable Zone}

To better understand the habitability of planets around K stars, we need to understand what the UV and X-ray flux incident on a planet in the habitable zone would be. Using the radii and effective temperature estimates of \citet{Baraffe2015}, we calculated the HZ distance as a function of age for a K star (0.8 M$_{\odot}$), an early-M star (0.4 M$_{\odot}$), and a late-M star (0.1 M$_{\odot}$) using Equation 3 of \citet{Kopparapu2013}:

\begin{equation}
    d = (\frac{R^2 \cdot T_{\rm eff}^4}{S_{\rm eff}})^{0.5} \text{AU},
\end{equation}

\noindent where the stellar radius $R$ and effective temperature $T_{\rm eff}$ are given in solar units and $S_{\rm eff}$ is the stellar photospheric flux at the HZ, which is strictly a function of the stellar temperature. The stellar flux at the HZ for each star was interpolated from the $S_{\rm eff}$ values given in \citet{kopparapu14} for temperatures from 2600 K - 7200 K. The HZ distances are shown in Figure \ref{fig:hz}a as a function of age. The opaque regions represent the conservative habitable zones and the translucent regions are the optimistic habitable zones. The size of the K-dwarf HZ is about twice as large as that of the M-stars and farther from the star by at least 0.4 AU.

We then calculated the range of NUV, FUV, and X-ray median values in the HZ at each age, as seen in Figures \ref{fig:hz}b, \ref{fig:hz}c, and \ref{fig:hz}d, respectively. In each of the plots, the K-star flux levels are below those of the M-stars by at least five times for the early-M stars and at least 50 times for the late-M stars. The initial rise in the NUV, FUV, and X-ray flux of all of the M stars is due to the HZ becoming closer-in to the star as the star ages, thus increasing the UV and X-ray flux in this region. For this reason, the UV and X-ray HZ flux of late-M stars is larger during the early stages of planet evolution than when the planet was initially formed. Conversely, the UV and X-ray HZ flux of K stars decreases from the time of planet formation, thus allowing more suitable conditions for the development and detectability of life, perhaps garnering K dwarfs the ``superhabitable" label.

\section{Implications and Conclusions}\label{sec:summary}

Because of their high-energy radiation environments, more frequent flaring, and tidally locked planets, M stars may not be the most suitable locations for finding habitable planets. Instead, if high UV emission is detrimental, then K stars offer a more favorable UV environment throughout the key evolutionary period of the planet and its atmosphere due to wider habitable zones and faster contraction onto the main sequence. 

Using \textit{GALEX} and \textit{ROSAT} photometry, we calculated the NUV and FUV excess (i.e. photosphere subtracted to yield the emission from the upper-atmosphere) flux densities and X-ray flux levels from young moving group members, clusters, and field stars ranging from 10 Myr to 5 Gyr. We find that the FUV and X-ray fluxes remain constant for $\sim$100 Myr before decreasing, as compared to M stars, which remain constant until $\sim$650 Myr \citep{Shkolnik2014, Schneider2018}. To determine the extent of the potential ``super-habitability" of K stars, we also analyzed these results within the habitable zones of K and M stars. We find that the UV and X-ray flux incident on a planet is 5 - 50 times lower than that of early-M stars  and  50 - 1000 times lower than  that  of  late-M  stars (depending on age). These UV and X-ray levels offer a much more suitable environment and a higher probability of planet habitability and the detection of any biomarkers. 

These results should be used to advise estimations of the extreme-UV, for which very little data currently exist. The EUV fluxes are some of the most effective measures of chemical alteration, as the photodissociation cross-sections are the largest for most molecules in this regime. The EUV flux is expected to follow the trends of both the FUV and X-ray emission. Therefore, we would expect the EUV to decline as early as $\sim$100 Myr.

The ratio of FUV to NUV is an important determinate in the probability of detecting a habitable planet. Larger ratios allow for longer lifetimes of biogenic gases such as as CH$_4$, N$_2$O, and CH$_3$Cl, leading to a higher probability of detection. However, this ratio also affects the abundances of abiotic molecules such as CO$_2$, O$_2$, and O$_3$ that may lead to false positive signatures. We find that this ratio is both mass- and age-dependent for all K spectral types. 

While these results are informative of median UV and X-ray fluxes incident on K star planets during the early stages of their evolution, this does not entail a higher probability of habitability in the case of all K dwarfs.

\vspace{0.5cm}

The authors would like to acknowledge support from the NASA Habitable Worlds grant NNX16AB62G. 
We wish to thank the anonymous referee for an insightful and helpful report. This work is based on observations made with the NASA \textit{Galaxy Evolution Explorer} and the \textit{R{\"O}ntgenSATellit}. \textit{GALEX} is operated for NASA by the California Institute of Technology under NASA contract NAS5-98034. This research utilized the public data from the second ROSAT All-Sky Survey (\url{https://heasarc.gsfc.nasa.gov/W3Browse/rosat/rass2rxs.html}). This work makes use of data products from the Two Micron All-Sky Survey, which is a joint project of the University of Massachusetts and the Infrared Processing and Analysis Center/California Institute of Technology, funded by the National Aeronautics and Space Administration and the National Science Foundation. This research has made use of the SIMBAD database, operated at CDS, Strasbourg, France. This work has made use of data from the European Space Agency (ESA) mission \textit{Gaia} (\url{https://www. cosmos.esa.int/gaia}), processed by the Gaia Data Processing and Analysis Consortium (DPAC, \url{https://www. cosmos.esa.int/web/gaia/dpac/consortium}).

\software{Astropy \, \citep{astropy:2018},\, Matplotlib \citep{matplotlib},\, Numpy \,\citep{numpy2}, \, Scipy \, \citep{scipy}, \,PHOENIX \,\citep{hauschildt1997, short2005}}

\bibliography{bibliography}

\begin{thebibliography}{}
\expandafter\ifx\csname natexlab\endcsname\relax\def\natexlab#1{#1}\fi

\bibitem[{Airapetian {et~al.}(2017)Airapetian, Glocer, Khazanov, Loyd, France,
  Sojka, Danchi, \& Liemohn}]{Airapetian2017}
Airapetian, V.~S., Glocer, A., Khazanov, G.~V., {et~al.} 2017, ApJ, 836, L3

\bibitem[{{Arney} {et~al.}(2018){Arney}, {Domagal-Goldman}, \&
  {Meadows}}]{Arney2018}
{Arney}, G., {Domagal-Goldman}, S.~D., \& {Meadows}, V. 2018, in American
  Astronomical Society Meeting Abstracts, Vol. 231, American Astronomical
  Society Meeting Abstracts \#231, 427.08

\bibitem[{Arney {et~al.}(2017)Arney, Meadows, Domagal-Goldman, Deming,
  Robinson, Tovar, Wolf, \& Schwieterman}]{Arney2017}
Arney, G.~N., Meadows, V.~S., Domagal-Goldman, S.~D., {et~al.} 2017, ApJ, 836,
  1, 49, 19 pp. (2017)., 836, arXiv:1702.02994

\bibitem[{{Baraffe} {et~al.}(2015){Baraffe}, {Homeier}, {Allard}, \&
  {Chabrier}}]{Baraffe2015}
{Baraffe}, I., {Homeier}, D., {Allard}, F., \& {Chabrier}, G. 2015, \aap, 577,
  A42

\bibitem[{{Barclay} {et~al.}(2018){Barclay}, {Pepper}, \&
  {Quintana}}]{barclay18}
{Barclay}, T., {Pepper}, J., \& {Quintana}, E.~V. 2018, \apjs, 239, 2

\bibitem[{Barnes \& Rory(2017)}]{Barnes2017}
Barnes, R., \& Rory. 2017, Celestial Mechanics and Dynamical Astronomy, 129, 4,
  pp.509-536, 129, 509

\bibitem[{Bell {et~al.}(2015)Bell, Mamajek, \& Naylor}]{Bell2015}
Bell, C. P.~M., Mamajek, E.~E., \& Naylor, T. 2015, MNRAS, 454, 1, p.593-614,
  454, 593

\bibitem[{{Boller} {et~al.}(2016){Boller}, {Freyberg}, {Tr{\"u}mper}, {Haberl},
  {Voges}, \& {Nandra}}]{boller}
{Boller}, T., {Freyberg}, M.~J., {Tr{\"u}mper}, J., {et~al.} 2016, \aap, 588,
  A103

\bibitem[{{Booth} {et~al.}(2017){Booth}, {Poppenhaeger}, {Watson}, {Silva
  Aguirre}, \& {Wolk}}]{booth17}
{Booth}, R.~S., {Poppenhaeger}, K., {Watson}, C.~A., {Silva Aguirre}, V., \&
  {Wolk}, S.~J. 2017, \mnras, 471, 1012

\bibitem[{Charbonneau \& Deming(2007)}]{Charbonneau2007}
Charbonneau, D., \& Deming, D. 2007, eprint arXiv:0706.1047, arXiv:0706.1047

\bibitem[{Checlair {et~al.}(2017)Checlair, Menou, \& Abbot}]{Checlair2017}
Checlair, J., Menou, K., \& Abbot, D.~S. 2017, ApJ, 845, 2, 132, 10 pp.
  (2017)., 845, arXiv:1705.08904

\bibitem[{Cockell {et~al.}(2016)Cockell, Bush, Bryce, Direito, Fox-Powell,
  Harrison, Lammer, Landenmark, Martin-Torres, Nicholson, Noack,
  O'Malley-James, Payler, Rushby, Samuels, Schwendner, Wadsworth, \&
  Zorzano}]{Cockell2016}
Cockell, C., Bush, T., Bryce, C., {et~al.} 2016, Astrobiology, 16, 89

\bibitem[{Cuntz \& Guinan(2016)}]{Cuntz2016}
Cuntz, M., \& Guinan, E.~F. 2016, ApJ, 827, 1, 79, 9 pp. (2016)., 827,
  arXiv:1606.09580

\bibitem[{{Domagal-Goldman} {et~al.}(2014){Domagal-Goldman}, {Segura},
  {Claire}, {Robinson}, \& {Meadows}}]{DG14}
{Domagal-Goldman}, S.~D., {Segura}, A., {Claire}, M.~W., {Robinson}, T.~D., \&
  {Meadows}, V.~S. 2014, \apj, 792, 90

\bibitem[{Findeisen {et~al.}(2011)Findeisen, Hillenbrand, \&
  Soderblom}]{Findeisen2011}
Findeisen, K., Hillenbrand, L., \& Soderblom, D. 2011, AJ, 142, 1, 23, 17 pp.
  (2011)., 142, arXiv:1105.1377

\bibitem[{{Fontenla} {et~al.}(2016){Fontenla}, {Linsky}, {Witbrod}, {France},
  {Buccino}, {Mauas}, {Vieytes}, \& {Walkowicz}}]{fontenla}
{Fontenla}, J.~M., {Linsky}, J.~L., {Witbrod}, J., {et~al.} 2016, \apj, 830,
  154

\bibitem[{{Gagn{\'e}} {et~al.}(2017){Gagn{\'e}}, {Faherty}, {Mamajek}, {Malo},
  {Doyon}, {Filippazzo}, {Weinberger}, {Donaldson}, {L{\'e}pine},
  {Lafreni{\`e}re}, {Artigau}, {Burgasser}, {Looper}, {Boucher}, {Beletsky},
  {Camnasio}, {Brunette}, \& {Arboit}}]{GagneTWA}
{Gagn{\'e}}, J., {Faherty}, J.~K., {Mamajek}, E.~E., {et~al.} 2017, \apjs, 228,
  18

\bibitem[{{Gaia Collaboration} {et~al.}(2018){Gaia Collaboration}, {Brown},
  {Vallenari}, {Prusti}, {de Bruijne}, {Babusiaux}, {Bailer-Jones}, {Biermann},
  {Evans}, {Eyer}, \& et~al.}]{gaiadr2}
{Gaia Collaboration}, {Brown}, A.~G.~A., {Vallenari}, A., {et~al.} 2018, \aap,
  616, A1

\bibitem[{{Goldman} {et~al.}(2013){Goldman}, {R{\"o}ser}, {Schilbach},
  {Magnier}, {Olczak}, {Henning}, {Juri{\'c}}, {Schlafly}, {Chen}, {Platais},
  {Burgett}, {Hodapp}, {Heasley}, {Kudritzki}, {Morgan}, {Price}, {Tonry}, \&
  {Wainscoat}}]{goldmanhyades}
{Goldman}, B., {R{\"o}ser}, S., {Schilbach}, E., {et~al.} 2013, \aap, 559, A43

\bibitem[{{Gondoin}(2017)}]{gondoin17}
{Gondoin}, P. 2017, \aap, 599, A122

\bibitem[{{Gondoin}(2018)}]{gondoin18}
---. 2018, ArXiv e-prints, arXiv:1808.01824

\bibitem[{{Harman} {et~al.}(2015){Harman}, {Schwieterman}, {Schottelkotte}, \&
  {Kasting}}]{harman15}
{Harman}, C.~E., {Schwieterman}, E.~W., {Schottelkotte}, J.~C., \& {Kasting},
  J.~F. 2015, \apj, 812, 137

\bibitem[{{Hauschildt} {et~al.}(1997){Hauschildt}, {Baron}, \&
  {Allard}}]{hauschildt1997}
{Hauschildt}, P.~H., {Baron}, E., \& {Allard}, F. 1997, \apj, 483, 390

\bibitem[{Heller \& Armstrong(2014)}]{Heller2014}
Heller, R., \& Armstrong, J. 2014, Astrobiology, Vol. 14, No. 1, p. 50-66., 14,
  50

\bibitem[{Hu {et~al.}(2012)Hu, Seager, \& Bains}]{Hu2012}
Hu, R., Seager, S., \& Bains, W. 2012, ApJ, 761, 2, 166, 29 pp. (2012)., 761,
  arXiv:1210.6885

\bibitem[{{Huang} {et~al.}(2018){Huang}, {Shporer}, {Dragomir}, {Fausnaugh},
  {Levine}, {Morgan}, {Nguyen}, {Ricker}, {Wall}, {Woods}, \&
  {Vanderspek}}]{huang18}
{Huang}, C.~X., {Shporer}, A., {Dragomir}, D., {et~al.} 2018, arXiv e-prints,
  arXiv:1807.11129

\bibitem[{{Hunter}(2007)}]{matplotlib}
{Hunter}, J.~D. 2007, Computing in Science and Engineering, 9, 90

\bibitem[{{Jackson} {et~al.}(2012){Jackson}, {Davis}, \&
  {Wheatley}}]{jackson12J}
{Jackson}, A.~P., {Davis}, T.~A., \& {Wheatley}, P.~J. 2012, \mnras, 422, 2024

\bibitem[{Jones {et~al.}(2001)Jones, Oliphant, Peterson, {et~al.}}]{scipy}
Jones, E., Oliphant, T., Peterson, P., {et~al.} 2001, {SciPy}: Open source
  scientific tools for {Python}, [Online; accessed <today>]

\bibitem[{{Jones} {et~al.}(2017){Jones}, {White}, {Boyajian}, {Schaefer},
  {Baines}, {Ireland}, {Quinn}, \& {CHARA Team}}]{Jones17}
{Jones}, J., {White}, R.~J., {Boyajian}, T.~S., {et~al.} 2017, in American
  Astronomical Society Meeting Abstracts, Vol. 229, American Astronomical
  Society Meeting Abstracts \#229, 131.05

\bibitem[{{Jones} {et~al.}(2015){Jones}, {White}, {Boyajian}, {Schaefer},
  {Baines}, {Ireland}, {Patience}, {ten Brummelaar}, {McAlister}, {Ridgway},
  {Sturmann}, {Sturmann}, {Turner}, {Farrington}, \& {Goldfinger}}]{Jones15b}
{Jones}, J., {White}, R.~J., {Boyajian}, T., {et~al.} 2015, \apj, 813, 58

\bibitem[{Kaltenegger(2017)}]{Kaltenegger2017}
Kaltenegger, L. 2017, Annual Review of A\&A, 55, 433

\bibitem[{Kasting {et~al.}(1993)Kasting, Whitmire, \& Reynolds}]{Kasting1993}
Kasting, J.~F., Whitmire, D.~P., \& Reynolds, R.~T. 1993, Icarus, 101, 108

\bibitem[{{Kislyakova} {et~al.}(2018){Kislyakova}, {Fossati}, {Johnstone},
  {Noack}, {L{\"u}ftinger}, {Zaitsev}, \& {Lammer}}]{Kislyakova2018}
{Kislyakova}, K.~G., {Fossati}, L., {Johnstone}, C.~P., {et~al.} 2018, \apj,
  858, 105

\bibitem[{{Kislyakova} {et~al.}(2017){Kislyakova}, {Noack}, {Johnstone},
  {Zaitsev}, {Fossati}, {Lammer}, {Khodachenko}, {Odert}, \&
  {G{\"u}del}}]{Kislyakova2017}
{Kislyakova}, K.~G., {Noack}, L., {Johnstone}, C.~P., {et~al.} 2017, Nature
  Astronomy, 1, 878

\bibitem[{Kopparapu {et~al.}(2013)Kopparapu, Ramirez, Kasting, Eymet, Robinson,
  Mahadevan, Terrien, Domagal-Goldman, Meadows, \& Deshpande}]{Kopparapu2013}
Kopparapu, R., Ramirez, R., Kasting, J.~F., {et~al.} 2013, ApJ, 765, 2, 131, 16
  pp. (2013)., 765, arXiv:1301.6674

\bibitem[{{Kopparapu} {et~al.}(2014){Kopparapu}, {Ramirez}, {SchottelKotte},
  {Kasting}, {Domagal-Goldman}, \& {Eymet}}]{kopparapu14}
{Kopparapu}, R.~K., {Ramirez}, R.~M., {SchottelKotte}, J., {et~al.} 2014,
  \apjl, 787, L29

\bibitem[{Koskinen {et~al.}(2010)Koskinen, Yelle, Lavvas, \&
  Lewis}]{Koskinen2010}
Koskinen, T.~T., Yelle, R.~V., Lavvas, P., \& Lewis, N.~K. 2010, ApJ, 723, 1,
  pp. 116-128 (2010)., 723, 116

\bibitem[{Kraus \& Hillenbrand(2007)}]{Kraus2007}
Kraus, A.~L., \& Hillenbrand, L.~A. 2007, AJ, 134, 6, pp. 2340-2352 (2007).,
  134, 2340

\bibitem[{{Kraus} {et~al.}(2014){Kraus}, {Shkolnik}, {Allers}, \&
  {Liu}}]{KrausTucHor}
{Kraus}, A.~L., {Shkolnik}, E.~L., {Allers}, K.~N., \& {Liu}, M.~C. 2014, \aj,
  147, 146

\bibitem[{Lammer {et~al.}(2007)Lammer, Lichtenegger, Kulikov, Grie{\ss}meier,
  Terada, Erkaev, Biernat, Khodachenko, Ribas, Penz, \& Selsis}]{Lammer2007}
Lammer, H., Lichtenegger, H.~I., Kulikov, Y.~N., {et~al.} 2007, Astrobiology,
  7, 185

\bibitem[{Lichtenegger {et~al.}(2010)Lichtenegger, Lammer, Grie{\ss}meier,
  Kulikov, von Paris, Hausleitner, Krauss, \& Rauer}]{Lichtenegger2010}
Lichtenegger, H., Lammer, H., Grie{\ss}meier, J.-M., {et~al.} 2010, Icarus,
  210, 1

\bibitem[{Lingam \& Loeb(2017)}]{Lingam2017}
Lingam, M., \& Loeb, A. 2017, eprint arXiv:1710.11134, arXiv:1710.11134

\bibitem[{{Linsky} \& {G{\"u}del}(2015)}]{Linsky2015}
{Linsky}, J.~L., \& {G{\"u}del}, M. 2015, in Astrophysics and Space Science
  Library, Vol. 411, Characterizing Stellar and Exoplanetary Environments, ed.
  H.~{Lammer} \& M.~{Khodachenko}, 3

\bibitem[{{MacGregor} {et~al.}(2018){MacGregor}, {Weinberger}, {Wilner},
  {Kowalski}, \& {Cranmer}}]{2018ApJ...855L...2M}
{MacGregor}, M.~A., {Weinberger}, A.~J., {Wilner}, D.~J., {Kowalski}, A.~F., \&
  {Cranmer}, S.~R. 2018, \apjl, 855, L2

\bibitem[{{Malo} {et~al.}(2013){Malo}, {Doyon}, {Lafreni{\`e}re}, {Artigau},
  {Gagn{\'e}}, {Baron}, \& {Riedel}}]{Malo2013}
{Malo}, L., {Doyon}, R., {Lafreni{\`e}re}, D., {et~al.} 2013, \apj, 762, 88

\bibitem[{Martin {et~al.}(2004)Martin, Fanson, Schiminovich, Morrissey,
  Friedman, Barlow, Conrow, Grange, Jelinsky, Milliard, Siegmund, Bianchi,
  Byun, Donas, Forster, Heckman, Lee, Madore, Malina, Neff, Rich, Small,
  Szalay, Wyder, Welsh, \& Wyder}]{Martin2004}
Martin, D.~C., Fanson, J., Schiminovich, D., {et~al.} 2004, ApJ, 619, 1, pp.
  L1-L6., 619, L1

\bibitem[{Miles \& Shkolnik(2017)}]{Miles2017}
Miles, B.~E., \& Shkolnik, E.~L. 2017, AJ, 154, 2, 67, 19 pp. (2017)., 154,
  arXiv:1705.03583

\bibitem[{{Montes} {et~al.}(2001){Montes}, {L{\'o}pez-Santiago}, {G{\'a}lvez},
  {Fern{\'a}ndez-Figueroa}, {De Castro}, \& {Cornide}}]{Montes2001}
{Montes}, D., {L{\'o}pez-Santiago}, J., {G{\'a}lvez}, M.~C., {et~al.} 2001,
  \mnras, 328, 45

\bibitem[{{Peacock} {et~al.}(2015){Peacock}, {Barman}, \& {Shkolnik}}]{peacock}
{Peacock}, S., {Barman}, T., \& {Shkolnik}, E. 2015, in AAS/Division for
  Planetary Sciences Meeting Abstracts, Vol.~47, AAS/Division for Planetary
  Sciences Meeting Abstracts \#47, 404.06

\bibitem[{{Pearson}(1895)}]{pearsonr2}
{Pearson}, K. 1895, Proceedings of the Royal Society of London Series I, 58,
  240

\bibitem[{{Pecaut} \& {Mamajek}(2013)}]{Pecaut2013}
{Pecaut}, M.~J., \& {Mamajek}, E.~E. 2013, \apjs, 208, 9

\bibitem[{Perryman {et~al.}(1997)Perryman, Brown, Lebreton, Gomez, Turon,
  de~Strobel, Mermilliod, Robichon, Kovalevsky, \& Crifo}]{Perryman1997}
Perryman, M. A.~C., Brown, A. G.~A., Lebreton, Y., {et~al.} 1997, A\&A, v.331,
  p.81-120 (1998), 331, 81

\bibitem[{Petigura {et~al.}(2013)Petigura, Howard, \& Marcy}]{Petigura2013}
Petigura, E.~A., Howard, A.~W., \& Marcy, G.~W. 2013, Proceedings of the
  National Academy of Sciences, vol. 110, 48, pp. 19273-19278, 110, 19273

\bibitem[{{Price-Whelan} {et~al.}(2018){Price-Whelan}, {Sip{\'{o}}cz},
  {G{\"u}nther}, {Lim}, {Crawford}, {Conseil}, {Shupe}, {Craig}, {Dencheva},
  {Ginsburg}, {VanderPlas}, {Bradley}, {P{'e}rez-Su{'a}rez}, {de Val-Borro},
  {Paper Contributors}, {Aldcroft}, {Cruz}, {Robitaille}, {Tollerud},
  {Coordination Committee}, {Ardelean}, {Babej}, {Bach}, {Bachetti}, {Bakanov},
  {Bamford}, {Barentsen}, {Barmby}, {Baumbach}, {Berry}, {Biscani}, {Boquien},
  {Bostroem}, {Bouma}, {Brammer}, {Bray}, {Breytenbach}, {Buddelmeijer},
  {Burke}, {Calderone}, {Cano Rodr{'i}guez}, {Cara}, {Cardoso}, {Cheedella},
  {Copin}, {Corrales}, {Crichton}, {D{ extquoteright}Avella}, {Deil},
  {Depagne}, {Dietrich}, {Donath}, {Droettboom}, {Earl}, {Erben}, {Fabbro},
  {Ferreira}, {Finethy}, {Fox}, {Garrison}, {Gibbons}, {Goldstein}, {Gommers},
  {Greco}, {Greenfield}, {Groener}, {Grollier}, {Hagen}, {Hirst}, {Homeier},
  {Horton}, {Hosseinzadeh}, {Hu}, {Hunkeler}, {Ivezi{'c}}, {Jain}, {Jenness},
  {Kanarek}, {Kendrew}, {Kern}, {Kerzendorf}, {Khvalko}, {King}, {Kirkby},
  {Kulkarni}, {Kumar}, {Lee}, {Lenz}, {Littlefair}, {Ma}, {Macleod},
  {Mastropietro}, {McCully}, {Montagnac}, {Morris}, {Mueller}, {Mumford},
  {Muna}, {Murphy}, {Nelson}, {Nguyen}, {Ninan}, {N{"o}the}, {Ogaz}, {Oh},
  {Parejko}, {Parley}, {Pascual}, {Patil}, {Patil}, {Plunkett}, {Prochaska},
  {Rastogi}, {Reddy Janga}, {Sabater}, {Sakurikar}, {Seifert}, {Sherbert},
  {Sherwood-Taylor}, {Shih}, {Sick}, {Silbiger}, {Singanamalla}, {Singer},
  {Sladen}, {Sooley}, {Sornarajah}, {Streicher}, {Teuben}, {Thomas},
  {Tremblay}, {Turner}, {Terr{'o}n}, {van Kerkwijk}, {de la Vega}, {Watkins},
  {Weaver}, {Whitmore}, {Woillez}, {Zabalza}, \& {Contributors}}]{astropy:2018}
{Price-Whelan}, A.~M., {Sip{\'{o}}cz}, B.~M., {G{\"u}nther}, H.~M., {et~al.}
  2018, AJ, 156, 123

\bibitem[{Ricker {et~al.}(2009)Ricker, Latham, Vanderspek, Ennico, Bakos,
  Brown, Burgasser, Charbonneau, Deming, Doty, Dunham, Elliot, Holman, Ida,
  Jenkins, Jernigan, Kawai, Laughlin, Lissauer, Martel, Sasselov, Schingler,
  Seager, Torres, Udry, Villasenor, Winn, \& Worden}]{Ricker2009}
Ricker, G.~R., Latham, D.~W., Vanderspek, R.~K., {et~al.} 2009, American
  Astronomical Society, AAS Meeting {\#}213, id.403.01; Bulletin of the
  American Astronomical Society, Vol. 41, p.193, 41, 193

\bibitem[{Ricker {et~al.}(2014)Ricker, Winn, Vanderspek, Latham, Bakos, Bean,
  Berta-Thompson, Brown, Buchhave, Butler, Butler, Chaplin, Charbonneau,
  Christensen-Dalsgaard, Clampin, Deming, Doty, {De Lee}, Dressing, Dunham,
  Endl, Fressin, Ge, Henning, Holman, Howard, Ida, Jenkins, Jernigan, Johnson,
  Kaltenegger, Kawai, Kjeldsen, Laughlin, Levine, Lin, Lissauer, MacQueen,
  Marcy, McCullough, Morton, Narita, Paegert, Palle, Pepe, Pepper, Quirrenbach,
  Rinehart, Sasselov, Sato, Seager, Sozzetti, Stassun, Sullivan, Szentgyorgyi,
  Torres, Udry, \& Villasenor}]{Ricker2014}
Ricker, G.~R., Winn, J.~N., Vanderspek, R., {et~al.} 2014, Proceedings of the
  SPIE, 9143, id. 914320 15 pp. (2014)., 9143, arXiv:1406.0151

\bibitem[{{Schaefer} \& {Fegley}(2010)}]{2010Icar..208..438S}
{Schaefer}, L., \& {Fegley}, B. 2010, \icarus, 208, 438

\bibitem[{{Schmitt} {et~al.}(1995){Schmitt}, {Fleming}, \&
  {Giampapa}}]{schmitt1995}
{Schmitt}, J.~H.~M.~M., {Fleming}, T.~A., \& {Giampapa}, M.~S. 1995, \apj, 450,
  392

\bibitem[{Schneider \& Shkolnik(2018)}]{Schneider2018}
Schneider, A.~C., \& Shkolnik, E.~L. 2018, AJ, 155, 3, 122, 16 pp. (2018).,
  155, arXiv:1801.06711

\bibitem[{Segura {et~al.}(2005)Segura, Kasting, Meadows, Cohen, Scalo, Crisp,
  Butler, \& Tinetti}]{Segura2005}
Segura, A., Kasting, J.~F., Meadows, V., {et~al.} 2005, Astrobiology, 5, 6, pp.
  706-725., 5, 706

\bibitem[{Segura {et~al.}(2003)Segura, Krelove, Kasting, Sommerlatt, Meadows,
  Crisp, Cohen, \& Mlawer}]{Segura2003}
Segura, A., Krelove, K., Kasting, J.~F., {et~al.} 2003, Astrobiology, 3, 689

\bibitem[{Segura {et~al.}(2010)Segura, Walkowicz, Meadows, Kasting, \&
  Hawley}]{Segura2010}
Segura, A., Walkowicz, L., Meadows, V., Kasting, J., \& Hawley, S. 2010,
  Astrobiology, 10, 7, pp. 751-771., 10, 751

\bibitem[{Shields {et~al.}(2016)Shields, Ballard, \& Johnson}]{Shields2016}
Shields, A.~L., Ballard, S., \& Johnson, J.~A. 2016, Physics Reports, Vol. 663,
  p. 1{\^{a}}€“38 (2016), 663, 1

\bibitem[{Shkolnik {et~al.}(2017)Shkolnik, Allers, Kraus, Liu, \&
  Flagg}]{Shkolnik2017}
Shkolnik, E.~L., Allers, K.~N., Kraus, A.~L., Liu, M.~C., \& Flagg, L. 2017,
  AJ, 154, 2, 69, 23 pp. (2017)., 154, arXiv:1706.04556

\bibitem[{Shkolnik \& Barman(2014)}]{Shkolnik2014}
Shkolnik, E.~L., \& Barman, T.~S. 2014, AJ, 148, 4, 64, 14 pp. (2014)., 148,
  arXiv:1407.1344

\bibitem[{{Short} \& {Hauschildt}(2005)}]{short2005}
{Short}, C.~I., \& {Hauschildt}, P.~H. 2005, \apj, 618, 926

\bibitem[{Spearman(1904)}]{Spearman1904}
Spearman, C. 1904, The American Journal of Psychology, 15, 72

\bibitem[{{Stelzer} {et~al.}(2013){Stelzer}, {Marino}, {Micela},
  {L{\'o}pez-Santiago}, \& {Liefke}}]{stezler13}
{Stelzer}, B., {Marino}, A., {Micela}, G., {L{\'o}pez-Santiago}, J., \&
  {Liefke}, C. 2013, \mnras, 431, 2063

\bibitem[{{Tian} {et~al.}(2014){Tian}, {France}, {Linsky}, {Mauas}, \&
  {Vieytes}}]{tian14}
{Tian}, F., {France}, K., {Linsky}, J.~L., {Mauas}, P.~J.~D., \& {Vieytes},
  M.~C. 2014, Earth and Planetary Science Letters, 385, 22

\bibitem[{{Truemper}(1982)}]{rosat}
{Truemper}, J. 1982, Advances in Space Research, 2, 241

\bibitem[{{van der Walt} {et~al.}(2011){van der Walt}, {Colbert}, \&
  {Varoquaux}}]{numpy2}
{van der Walt}, S., {Colbert}, S.~C., \& {Varoquaux}, G. 2011, Computing in
  Science and Engineering, 13, 22

\bibitem[{{van Leeuwen}(2009)}]{vanleeuwen}
{van Leeuwen}, F. 2009, \aap, 497, 209

\bibitem[{{Voges} {et~al.}(1999){Voges}, {Aschenbach}, {Boller},
  {Br{\"a}uninger}, {Briel}, {Burkert}, {Dennerl}, {Englhauser}, {Gruber},
  {Haberl}, {Hartner}, {Hasinger}, {K{\"u}rster}, {Pfeffermann}, {Pietsch},
  {Predehl}, {Rosso}, {Schmitt}, {Tr{\"u}mper}, \& {Zimmermann}}]{voges1999}
{Voges}, W., {Aschenbach}, B., {Boller}, T., {et~al.} 1999, \aap, 349, 389

\bibitem[{{Wang} {et~al.}(1995){Wang}, {Chen}, {Zhao}, \& {Jiang}}]{wang1995}
{Wang}, J.~J., {Chen}, L., {Zhao}, J.~H., \& {Jiang}, P.~F. 1995, \aaps, 113,
  419

\bibitem[{Zerkle {et~al.}(2012)Zerkle, Claire, Domagal-Goldman, Farquhar, \&
  Poulton}]{Zerkle2012}
Zerkle, A.~L., Claire, M.~W., Domagal-Goldman, S.~D., Farquhar, J., \& Poulton,
  S.~W. 2012, Nature Geoscience, 5, 359

\end{thebibliography}
\begin{longrotatetable}
\begin{deluxetable}{c c c c c c c c c c c c c}
\tabletypesize{\tiny}
\tablecolumns{12}
\tablewidth{5\textwidth}
\centering
\tablecaption{Observed flux densities $F$, fluxes $f$, and model photospheric flux densities $\sigma$ of the stars used in this analysis. \label{tab:data}}
\tablehead{
\colhead{Name\tablenotemark{a}} & \colhead{RA} & \colhead{Dec} & \colhead{Mass} & \colhead{Dist.}   & \colhead{F$_J$} & \colhead{$\sigma_J$} &  \colhead{F$_{NUV}$}   & \colhead{$\sigma_{NUV}$} &  \colhead{F$_{FUV}$}   & \colhead{$\sigma_{FUV}$} & \colhead{f$_{X}$}\\
\colhead{ } & \colhead{2MASS} & \colhead{2MASS} & \colhead{[M$_\odot$]} & \colhead{[pc]}   & \colhead{[$\mu$Jy]} & \colhead{[$\mu$Jy]} &  \colhead{[$\mu$Jy]}   & \colhead{[$\mu$Jy]} &  \colhead{[$\mu$Jy]}   & \colhead{[$\mu$Jy]} & \colhead{[erg cm$^{-2}$ s$^{-1}$]}\\
}

\startdata
\hline
\hline
    \multicolumn{12}{c}{TW Hydra, 10 Myr}\\
\hline
\hline
J11015191-3442170 & 11 01 51.9052 & -34 42 17.033 & 0.61 & 60.09 & 1.12E+07 & 6.13E+06 & > 9.29E+03  & 6.79E+01 & 9.20E+04 $\pm$ 1.85E+03 & 5.13E-04 & 2.36E-10 $\pm$ 1.70E-11 \\
J11482422-3728491 & 11 48 24.2227 & -37 28 49.114 & 0.75 & 76.38 & 3.12E+07 & 2.03E+07 & 1.44E+04   $\pm$ 6.60E+02 & 9.29E+01 & 1.55E+03  $\pm$ 4.04E+02 & 6.47E-04 & 1.65E-10 $\pm$ 2.56E-11 \\
J12153072-3948426 & 12 15 30.7230 & -39 48 42.594 & 0.68 & 53.11 & 2.43E+07 & 1.80E+07 & 4.14E+03   $\pm$ 2.41E+02 & 7.05E+01 & ---  $\pm$ --- & --- & 8.16E-11 $\pm$ 1.32E-11 \\
\hline
\hline
    \multicolumn{12}{c}{Beta Pictoris, 23 Myr}\\
\hline
\hline
J00233468+2014282 & 00 23 34.6661 & +20 14 28.637 & 0.71 & 63.15 & 3.53E+07 & 1.18E+07 & 7.73E+03 $\pm$ 4.20E+02 & 6.70E+01 & --- $\pm$ --- & --- & 6.59E-11 $\pm$ 8.22E-12 \\
J01373940+1835332 & 01 37 39.4099 & +18 35 33.259 & 0.89 & 52.15 & 2.71E+07 & 3.24E+07 & > 3.75E+03   & 3.07E+04 & 4.76E+04  $\pm$ 1.02E+03 & 7.19E-03 & 1.62E-10 $\pm$ 9.06E-12 \\
J02272924+3058246 & 02 27 29.2543 & +30 58 24.596 & 0.74 & 41.05 & 1.91E+07 & 1.29E+07 & 5.38E+03   $\pm$ 1.25E+02 & 8.24E+01 & 5.29E+02  $\pm$ 6.08E+01 & 1.10E-03 & 6.49E-11 $\pm$ 4.98E-12 \\
\enddata
\tablenotetext{a}{Names starting with J are 2MASS objects.}

\end{deluxetable}
\end{longrotatetable}
\normalsize

\end{document}